\documentclass[aps,prl,preprint,groupedaddress]{revtex4-2}
\usepackage{graphicx}
\usepackage{xcolor}
\usepackage{epstopdf}
\usepackage{color,soul}
\usepackage[utf8]{inputenc}
\usepackage{hyperref}
\usepackage{lineno}
\usepackage{amsmath, bm}
\usepackage{mathtools}
\usepackage{upgreek}
\usepackage[normalem]{ulem}
\usepackage{lineno}
\begin{document}

\title{Transverse and Longitudinal Magnetothermopower Promoted by Ambipolar Effect in Half-Heusler Topological Materials}
\author{Orest Pavlosiuk$^{1,*}$} 
\author{Marcin Matusiak$^{1,2}$} 
\author{Andrzej Ptok$^3$}
\author{Piotr Wiśniewski$^1$} 
\author{Dariusz Kaczorowski$^{1,*}$} 
\affiliation{$^1$~Institute of Low Temperature and Structure Research, Polish Academy of Sciences, Ok\'{o}lna 2, 50-422 Wroc{\l}aw, Poland\\
$^2$~International Research Centre MagTop, Institute of Physics, Polish Academy of Sciences, Aleja Lotnikow 32/46, PL-02668 Warsaw, Poland\\	 
$^3$~Institute of Nuclear Physics, Polish Academy of Sciences, W. E. Radzikowskiego 152, PL-31342 Krak\'{o}w, Poland\\
$^*$~Corresponding authors: d.kaczorowski@intibs.pl, o.pavlosiuk@intibs.pl}

\begin{abstract}
Topologically trivial and non-trivial semimetals with a high degree of carrier compensation are well known for demonstrating large transverse magnetothermopower ($S_{yx}$). 
However, in such systems, the longitudinal magnetothermopower ($S_{xx}$) is typically suppressed due to nearly perfect electron-hole compensation.
Here, we show that the half-Heusler topological semimetal DyPtBi exhibits simultaneously large $S_{xx}$ and $S_{yx}$ magnetothermopowers, defying this conventional trade-off. 
In $B=14$\,T, thermopower of DyPtBi reaches peak values of $S_{xx}=131\,\upmu\rm{V/K}$ at $T=149$\,K and $S_{yx}=-297\,\upmu\rm{V/K}$ at $T=200$\,K, and transverse component remains significantly large even at $290$\,K ($S_{yx}=-213\,\upmu\rm{V/K}$). 
Remarkably, at $T=290$\,K and in relatively weak magnetic field of $1$\,T, both relevant for practical applications, DyPtBi shows $S_{yx}=-18\,\upmu\rm{V/K}$, which is one of the largest values reported under such conditions. 
The large transverse thermopower originates from an ambipolar effect associated with thermal excitation occurring in zero-gap semiconductors. 
Due to the imperfect electron-hole compensation, an intrinsic asymmetry between hole- and electron-type carriers enables pronounced values of both $S_{xx}$ and $S_{yx}$, resulting in high effective thermopower ($S_{xx}+|S_{yx}|=379\,\upmu\rm{V/K}$) in DyPtBi at 200\,K. 
A comparative analysis with DyPdBi, another half-Heusler material that demonstrates large $S_{xx}=123\,\upmu\rm{V/K}$ but small $S_{yx}=-16\,\upmu\rm{V/K}$ (both values obtained at $T=293$\,K and $B=14$\,T), highlights the critical role of band structure and compensation tuning. 
These findings underscore the potential of chemical doping and band engineering in rare-earth-based half-Heusler materials for optimizing both transverse and longitudinal thermoelectric properties.
\end{abstract}
\maketitle

\section{Introduction}

Thermoelectric effects rely on a material ability to convert heat into voltage and vice versa.\cite{Bell2008a} 
Beyond fundamental interest, thermoelectric effects are crucial for practical applications in energy harvesting and solid-state cooling.
Despite thermoelectric materials are already used in real-world devices, significant research efforts are still devoted to enhance their thermoelectric performance. 
Two main metrics quantify such performance: the power factor (PF=$S^2/\rho$), which determines the output power, and the thermoelectric figure of merit ($zT=S^2T/(\rho\kappa)$), which determines the efficiency of energy conversion.\cite{Yan2022b} 
In the above formulas, $S$ is the Seebeck coefficient, $\rho$ the electrical resistivity, $\kappa$  the thermal conductivity, and $T$ temperature. 
Strategies to improve PF and $zT$ primarily focus on optimizing $S$, $\rho$ and $\kappa$. 
One promising way is the application of a magnetic field, which can significantly impact Seebeck coefficient (longitudinal thermopower).  
Additionally, magnetic field gives rise to the Nernst effect or so-called Nernst thermopower (transverse thermopower),\cite{V.Ettingshausen1886} which has been observed large in several topological semimetals, such as ZrTe$_5$,\cite{Wang2021t} WTe$_2$,\cite{Pan2022} NdP,\cite{Fu2018b,Scott2023} NbSb$_2$,\cite{Li2022d} TaAs$_2$,\cite{Hu2025} Mg$_3$Bi$_2$\cite{Feng2022a} and NbAs$_2$.\cite{Wu2024}
It is worth noting that the maximum Nernst thermopower values in all these semimetals have been observed at temperatures well below room temperature, which disqualifies them from applications under ambient conditions. 
In many of these materials, their unusually large Nernst effect has been attributed to nearly-perfect electron-hole compensation and ultrahigh mobility of charge carriers in topologically non-trivial bands.\cite{Wang2021t,Pan2022,Li2022d} 
Another contributing mechanism involves reaching the extreme quantum limit,\cite{Hu2025} which, along with Dirac band contributions\cite{Scott2023} and Zeeman splitting,\cite{Pan2025} can lead to a large magneto-Seebeck effect.
Importantly, some of these mechanism contribute to both magneto-thermoelectric effects and as a result some materials may show large magneto-Seebeck and Nernst effect simultaneously.\cite{Scott2023,Hu2025} 
However, ideal electron-hole compensation typically suppresses longitudinal thermopower while enhancing the Nernst effect.\cite{Chen2021l} 
In general, assuming equal carrier mobilities, a fundamental trade-off exists between the transverse and longitudinal thermoelectric responses in uncompensated semimetals, enhancement of one typically suppresses the other. 
This suggests that optimizing semimetals for thermoelectric performance may seem futile, as depending on the compensation degree one typically obtains either a strong response in one component and a weak one in the other, or two mediocre values.
However, if compensation is not perfect, or if mobilities of hole-type and electron-type carriers are different (i.e., finite electron-hole asymmetry exists), both longitudinal and transverse thermopowers can attain substantial values.\cite{Feng2022a,Gui2024,Feng2021} 
Interestingly, a recent device concept has demonstrated that simultaneous using of both thermoelectric responses can significantly enhance the effective power factor.\cite{Scott2023} 
This highlights the importance of identifying new materials, which can demonstrate large values of both longitudinal and transverse thermopowers due to the ambipolar effect (i.e., tuning a balance between electron and hole-type of carriers). 
Beyond semimetals, the ambipolar effect can also emerge in narrow gap-semiconductors, where the thermal broadening of the electronic distribution enables significant thermal excitation of both carrier types even at low temperatures.\cite{Delves1965}    
Nevertheless, there are other mechanisms that can facilitate enhancement of both transverse and longitudinal thermopower. 
These include Dirac-band features,\cite{Scott2023} and multi-pocket synergy between topologically trivial bands (responsible for large Nernst effect due to compensation mechanism) and topologically non-trivial bands (responsible for large magneto-Seebeck effect).\cite{Hu2025}
Implementing topologically non-trivial bands into uncompensated semimetals or narrow-band semiconductors can therefore enable simultaneous increase in both longitudinal and transverse thermopowers. 
Half-Heusler phases with the chemical composition $RE$PtBi (where $RE$ is a rare-earth element) are particularly promising in this regard. 
These materials are semimetals, featuring topologically non-trivial band structures that are induced by external magnetic field due to the strong Zeeman splitting\cite{Hirschberger2016a} and demonstrate some degree of carrier compensation.\cite{Schindler2018b,Luptbi2015,Pavlosiuk2020}
A large magneto-thermoelectric response has recently been reported in TbPtBi,\cite{Wang2022h} prompting us to investigate DyPtBi in the same context. 
Additionally, we extend our interest to DyPdBi, a member of half-Heusler $RE$PdBi group, which, to the best of our knowledge, has not been explored in therms magneto-thermoelectric properties. 
Notably, $RE$PdBi phases are also proposed to host magnetic-field-induced Weyl nodes.\cite{Pavlosiuk2019}
Herein, through a comparative study of DyPdBi and DyPtBi, including electronic structure calculations, measurements of magneto-transport properties and magneto-thermopowers, we demonstrate that the ambipolar effect plays a crucial role in governing their magneto-thermoelectric behavior. 
Both materials demonstrate non-ideal carrier compensation of different degree and finite asymmetry of hole and electron bands. 
As a result, DyPtBi demonstrates large and strongly magnetic field-dependent transverse and longitudinal magneto-thermopowers, while DyPdBi demonstrates a much smaller Nernst thermopower but a comparably large longitudinal magneto-thermopower, that is less sensitive to magnetic field. 
Our findings show that in zero-gap semiconductors, large transverse and longitudinal magneto-thermoelectric responses can persist even up to room temperature. 
This opens new opportunities for enhancing the performance of thermoelectric devices through band structure engineering in ambipolar systems.    

\section{Results and discussion}

\subsection{Electronic structure calculations}

\begin{figure}[h]
	\includegraphics[width=0.95\textwidth]{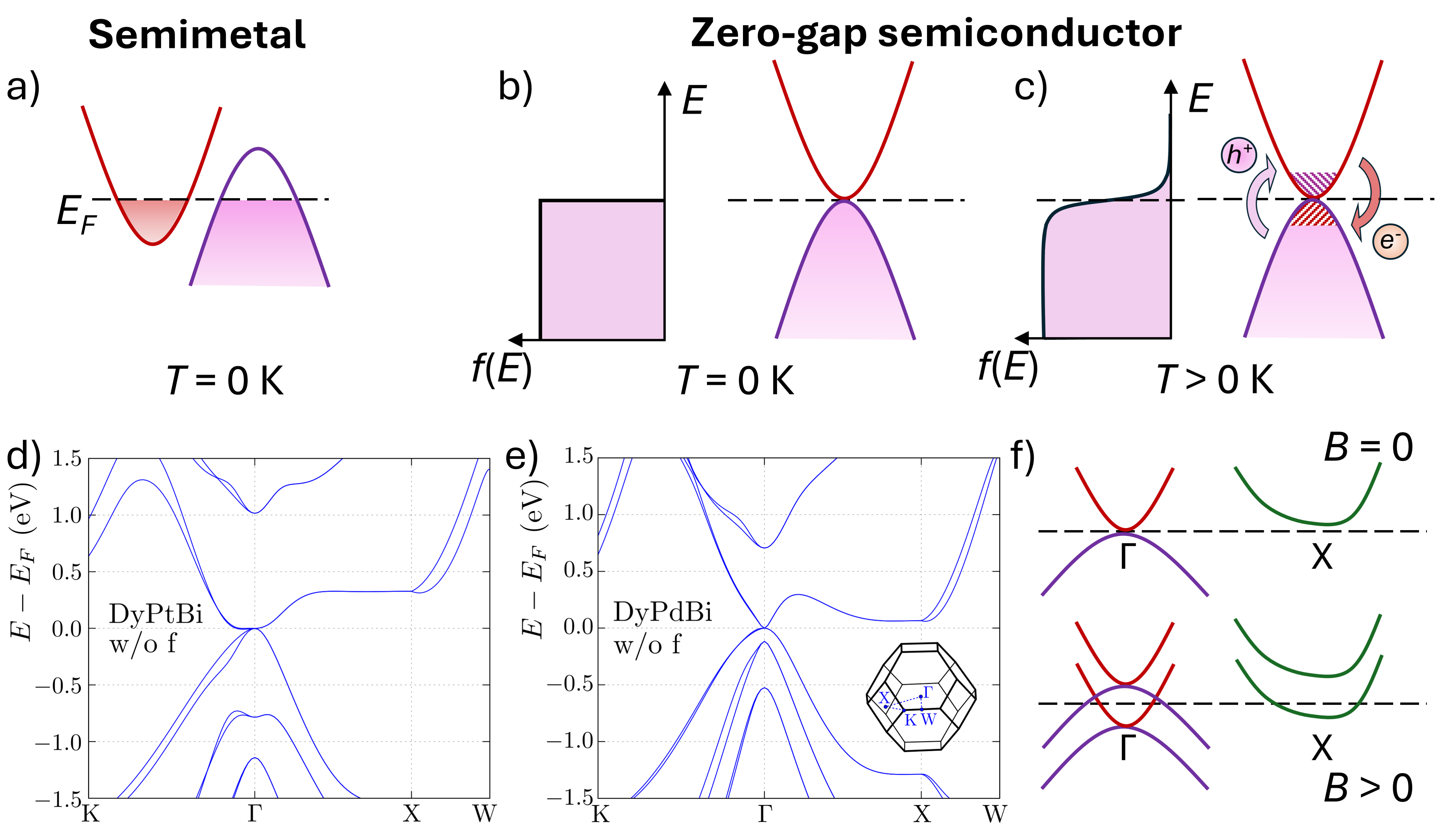}
	\caption{Schematics of the electronic structure for a semimetal (a) and a zero-gap semiconductor (b) at $T$=0\,K. The corresponding Fermi-Dirac distribution is also shown in panel (b). (c) Schematic representation of electronic structure and Fermi-Dirac distribution in a zero-gap semiconductor at $T>0$\,K, where thermal broadening enables the thermal excitation of carriers across the band touching point, leading to the multiband conduction involving both electons and holes. Panels (d) and (e) show the calculated electronic band structure of DyPtBi and DyPdBi, respectively, obtained for Dy $f$ states treated as core states. Band structures are plotted along high-symmetry directions of the Brillouin zone shown in the inset to panel (e). (f) Schematic illustration of the electronic band structure of DyPdBi in zero magnetic field and in finite magnetic field ($B>0$). The Zeeman effect causes band splitting, resulting in the emergence of multi-band transport.      
	\label{Electr_str}}
\end{figure}

Electronic structure of DyPtBi and DyPdBi, calculated without taking into account magnetic structure, is shown in Fig.\,\ref{Electr_str}d and Fig.\,\ref{Electr_str}e, respectively.   
Despite being isostructural and chemically similar, DyPdBi and DyPtBi demonstrate considerable differences in their electronic structure. 
As is evidenced from Fig.\,\ref{Electr_str}e, DyPdBi has a zero density of states at the Fermi level, as its Fermi level lies precisely at the zero-band gap. 
In contrast, in DyPtBi the Fermi level intersects tiny electron-like band along the $\Gamma\!-\!K$ high-symmetry line. 
Moreover, similar to DyPdBi, DyPtBi also shows a band touching point at $\Gamma$, where electron and hole-like pockets meet at the Fermi level, suggesting that it also retains features of a zero-gap semiconductor.
However, our calculations revealed the degree of asymmetry between the valence and conduction bands located around $\Gamma$ point is notably larger in DyPdBi than in DyPtBi.  
In zero-gap semiconductors, the thermal broadening plays important role in the physical properties. 
In case of DyPtBi and DyPdBi, the thermal broadening, which leads to thermal excitation of carriers, is especially relevant for bands at the $\Gamma$ point, where a zero band gap facilitates the activation of both electron- and hole-type carriers even at relatively low temperatures (see Fig.\,\ref{Electr_str}c). 
The thermally driven carrier activation enhances multi-band contributions to charge transport and thermopower, further influencing the overall thermoelectric response.\cite{Delves1965}

Another key feature of zero-gap semiconductors and semimetals is their pronounced susceptibility to Zeeman splitting effect, especially in materials with strong spin-orbit coupling (SOC). 
In this study, the compounds investigated have large SOC, as both containing of Bi, which has the heaviest nucleus among non-radioactive elements. 
It has been proposed that in GdPtBi Zeeman splitting is responsible for the appearance of Weyl nodes\cite{Hirschberger2016a} and this mechanism has then been found applicable to other $RE$PtBi compounds\cite{Pavlosiuk2020,Chen2020a}. 
Interestingly, in other studies, the exchange splitting has been reported to be the reason of Weyl nodes formation in GdPtBi\cite{Shekhar2018} and this idea has been then extended to $RE$AuSn systems.\cite{Ueda2025,Ueda2023} 
However, apart from the formation of Weyl nodes close to the $\Gamma$ point, Zeeman or exchange splitting can lead to other important modifications of the electronic structure and thus physical properties. 
In both compounds, Zeeman effect can lead to multi-band conductivity, as both electron- and hole-type bands at the $\Gamma$ point can be intersected by the Fermi level. 
Moreover, in DyPdBi, along the high-symmetry line $\Gamma\!-\!X$, the heavy electron-like band lies in close vicinity to the Fermi level (see Fig.\ref{Electr_str}e), and Zeeman splitting may force this band to intersect the Fermi level under magnetic field (see scheme Fig.\,\ref{Electr_str}f).  

It should also be kept in mind that in low-carrier-density systems, including half-Heusler phases $RE$PtBi and $RE$PdBi, the position of the Fermi level is strongly affected by the intrinsic crystal defects. 
Therefore, the results of electronic structure calculations must be interpreted with care. 
Previous studies have shown that in several half-Heusler compounds, including YbPtBi, LuPdBi, GdPtBi and LuPtBi, the experimentally determined position of Fermi level can deviate significantly from theoretical predictions and does not necessarily lie at the center of nominal  zero-band-gap.\cite{Ishihara2021a,Pavlosiuk2025,Guo2018,Liu2011a,Liu2016a}

In addition, we found that antiferromagnetic ordering, occurring below $3.3$\,K in DyPtBi\cite{Canfield1991} and $3.5$\,K in DyPdBi,\cite{Gofryk2005} significantly changes the electronic structure (see Supporting Information and Fig.\,S1).

\subsection{Electrical resistivity and thermoelectric power in zero magnetic field}

\begin{figure}[h]
	\includegraphics[width=0.49\textwidth]{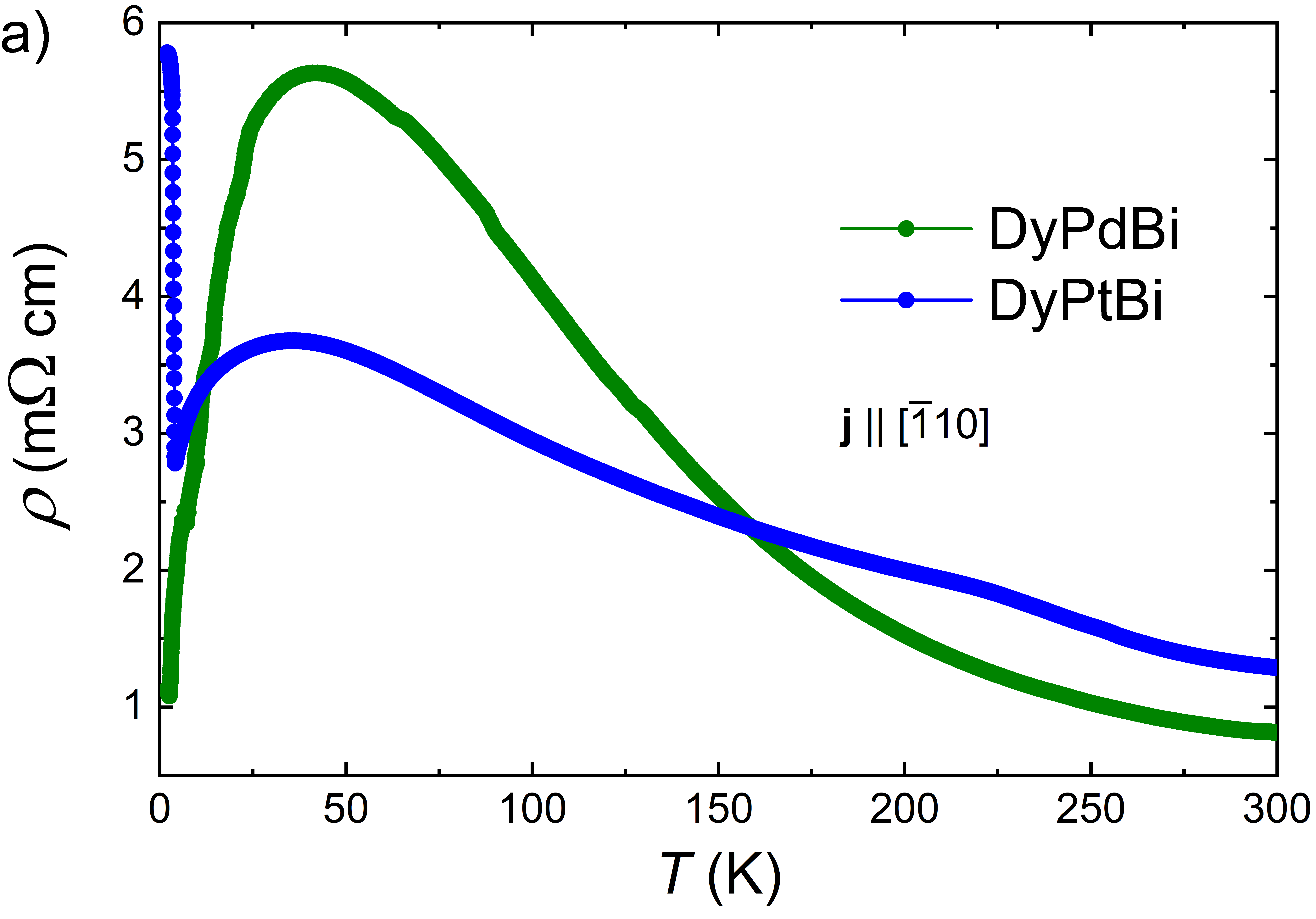}
	\includegraphics[width=0.49\textwidth]{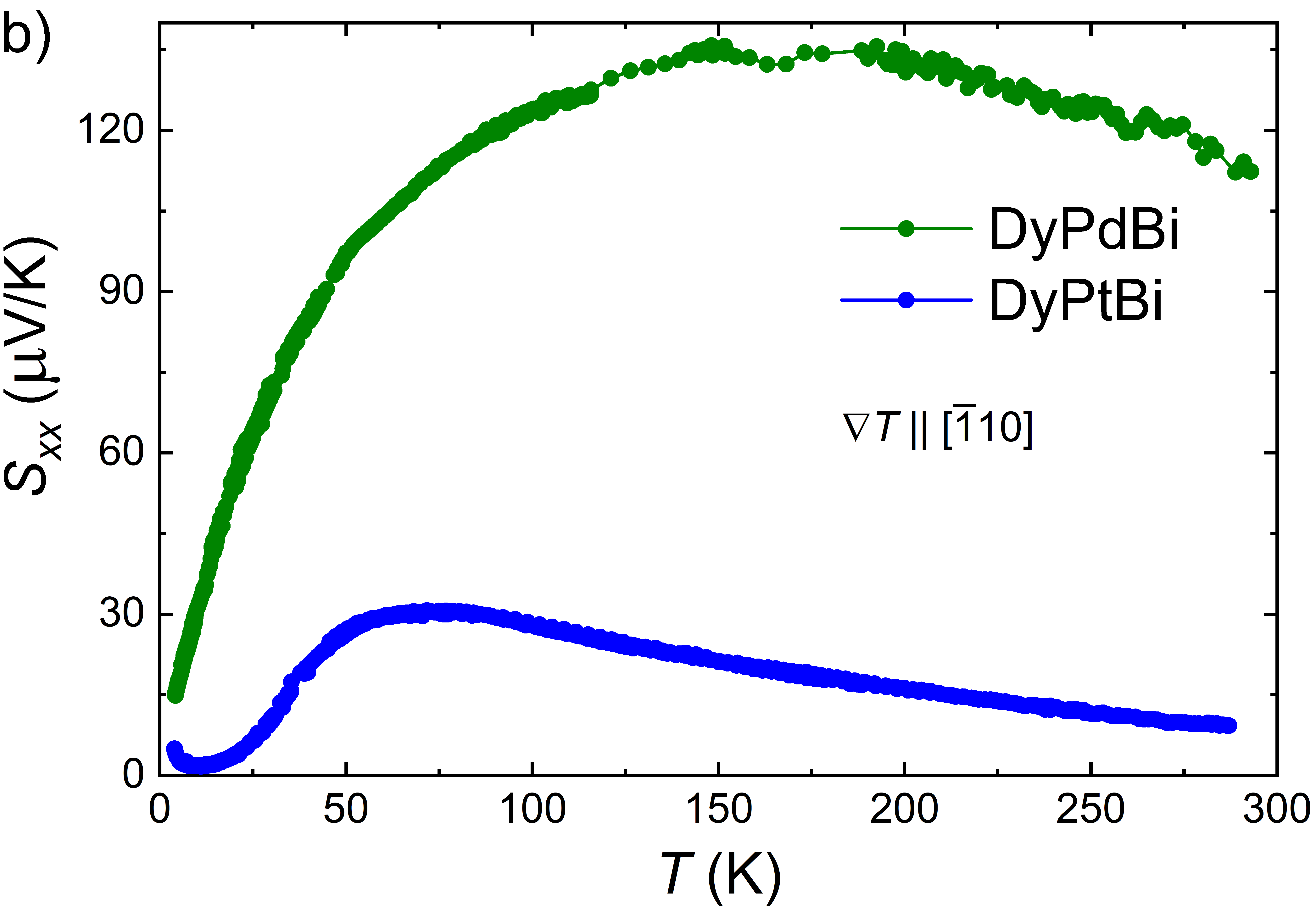}
	\caption{(a) Temperature dependence of the electrical resistivity of DyPtBi and DyPdBi, for both compounds measured with electrical current applied along [$\bar{1}10$] crystallographic direction. (b) Temperature dependence of the longitudinal thermopower of DyPtBi and DyPdBi, for both compounds measured with temperature gradient applied along [$\bar{1}10$] crystallographic direction.
		\label{rho_Sxx}}
\end{figure}

The temperature dependences of electrical resistivity ($\rho$) and longitudinal thermopower ($S_{xx}$) of DyPtBi and DyPdBi are shown in Fig.\,\ref{rho_Sxx}. 
In the paramagnetic state, $\rho(T)$ of both compounds displays a similar trend (see Fig.\,\ref{rho_Sxx}a). 
In the high-temperature range, $\rho(T)$ demonstrates semiconducting-like behavior, which evolves into a metallic-like response at lower temperatures. 
The overall behavior of $\rho(T)$ of both materials is typical of zero-gap semiconductors or semimetals,\cite{Dornhause1983} and align with the observations reported previously for these and other $RE$PtBi\cite{Hirschberger2016a,Chen2021h,Zhang2020k,Chen2020a,Chen2021b,Pavlosiuk2025} and $RE$PdBi compounds.\cite{Pavlosiuk2019,Zhu2023b,Pavlosiuk2016a,Pavlosiuk2016c}
Moreover, the magnitude of resistivity of both compounds is of the same order of magnitude.
For a more detailed description of $\rho(T)$, refer to Supporting Information.

Similarly to $\rho(T)$, $S_{xx}(T)$ of DyPdBi and DyPtBi share some similar features (see Fig.\,\ref{rho_Sxx}b).
Firstly, for both compounds, $S_{xx}$ is positive throughout the entire covered temperature range, suggesting that hole-type carriers dominate the electron transport. 
Secondly, within the high temperature range, $S_{xx}$ of both materials increases with decreasing $T$, thereafter reaches local maxima and starts to decrease.
In general, $S_{xx}(T)$ dependences observed by us for DyPtBi is qualitatively very similar to that observed in these material in Refs.\cite{Mun2016a,Gofryk2011} as well as other $RE$PtBi and $RE$PdBi compounds.\cite{Kim2001,Mun2016a,Gofryk2011,Kaczorowski2005,Gofryk2005,Mukhopadhyay2019}
The overall shape of $S_{xx}(T)$ curve for both materials points to a complex electronic structure, comprising both electron- and hole-like bands.\cite{Gofryk2011} 
This interpretation is consistent with the findings of electronic structure studies of $RE$PdBi and $RE$PtBi compounds\cite{Zhu2023b,Suzuki2016,Wang2022h} as well as our own electronic structure calculations.
However, $S_{xx}$ of DyPdBi is substantially higher than that of DyPtBi, which may be attributed to the differences in the degree of carrier compensation and/or differences in the mobilities of hole- and electron-type carriers in two materials (see Supporting Information for details.)
It should be noted that a similar behavior of $\rho(T)$ and $S_{xx}(T)$ has been observed in the narrow-gap semiconductor ZrTe$_5$, where it was attributed to the ambipolar conduction.\cite{Shahi2018}

\subsection{Longitudinal magneto-thermoelectric effect}

\begin{figure}[h]
	\includegraphics[width=0.49\textwidth]{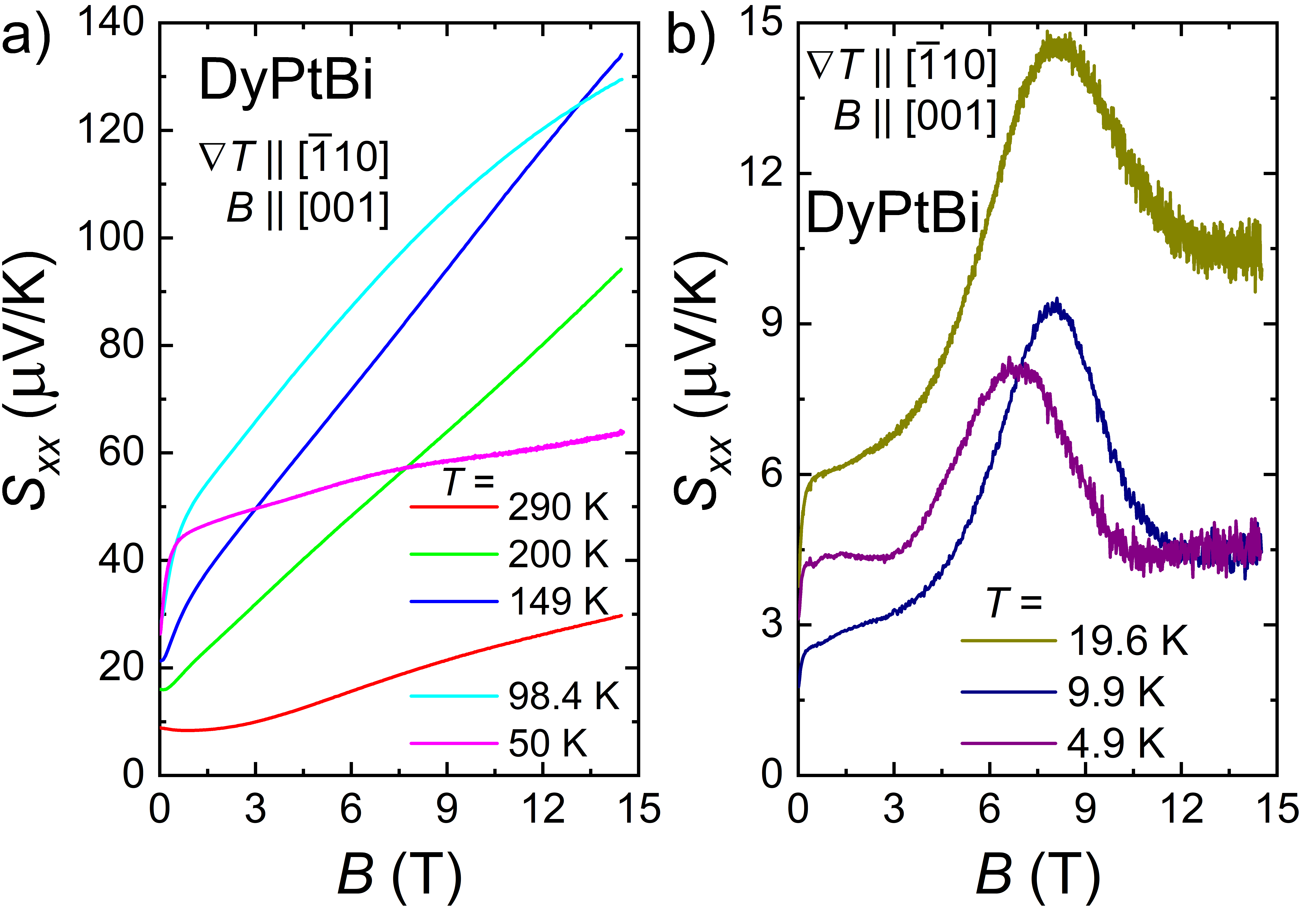}
	\includegraphics[width=0.49\textwidth]{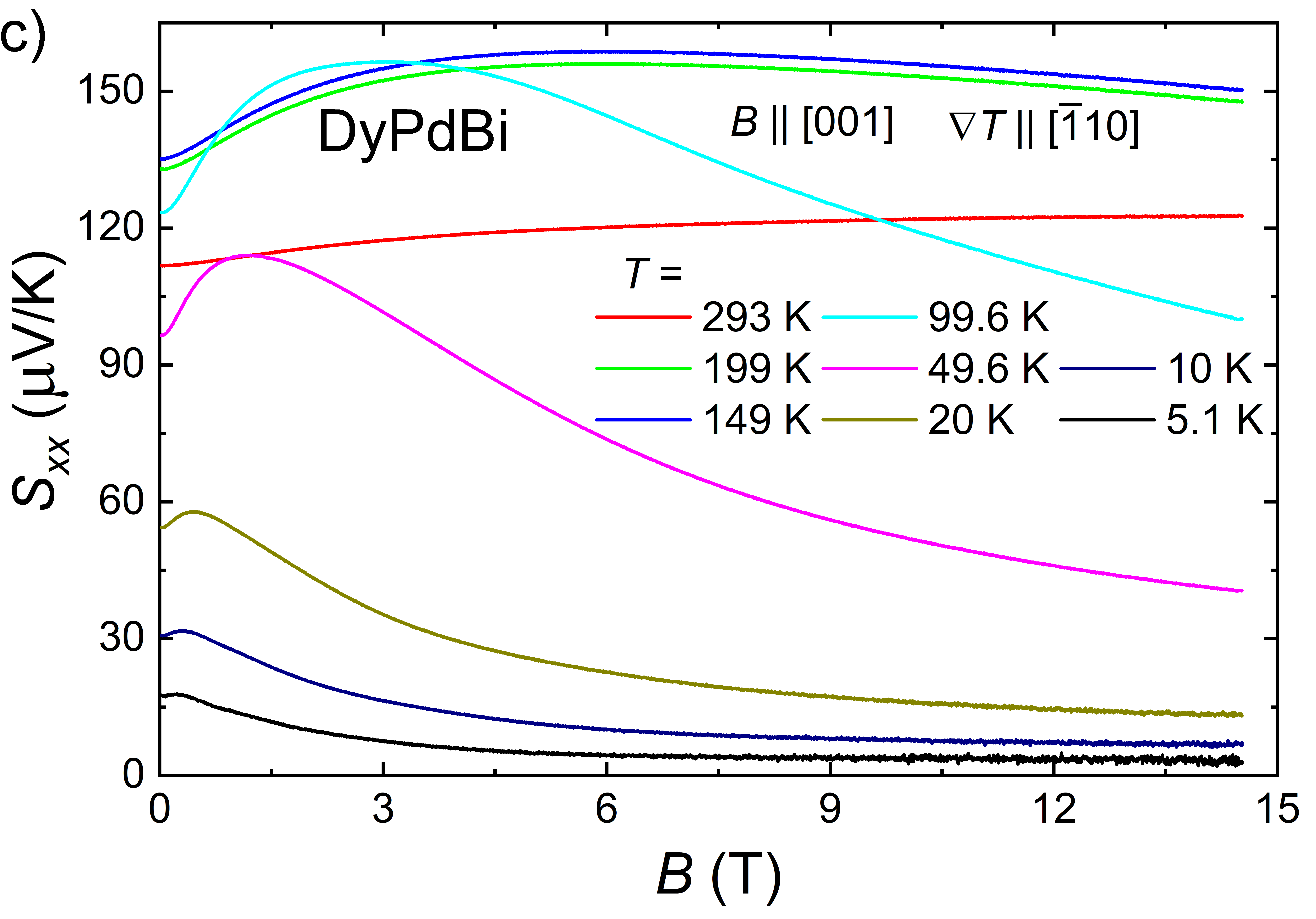}
	\includegraphics[width=0.49\textwidth]{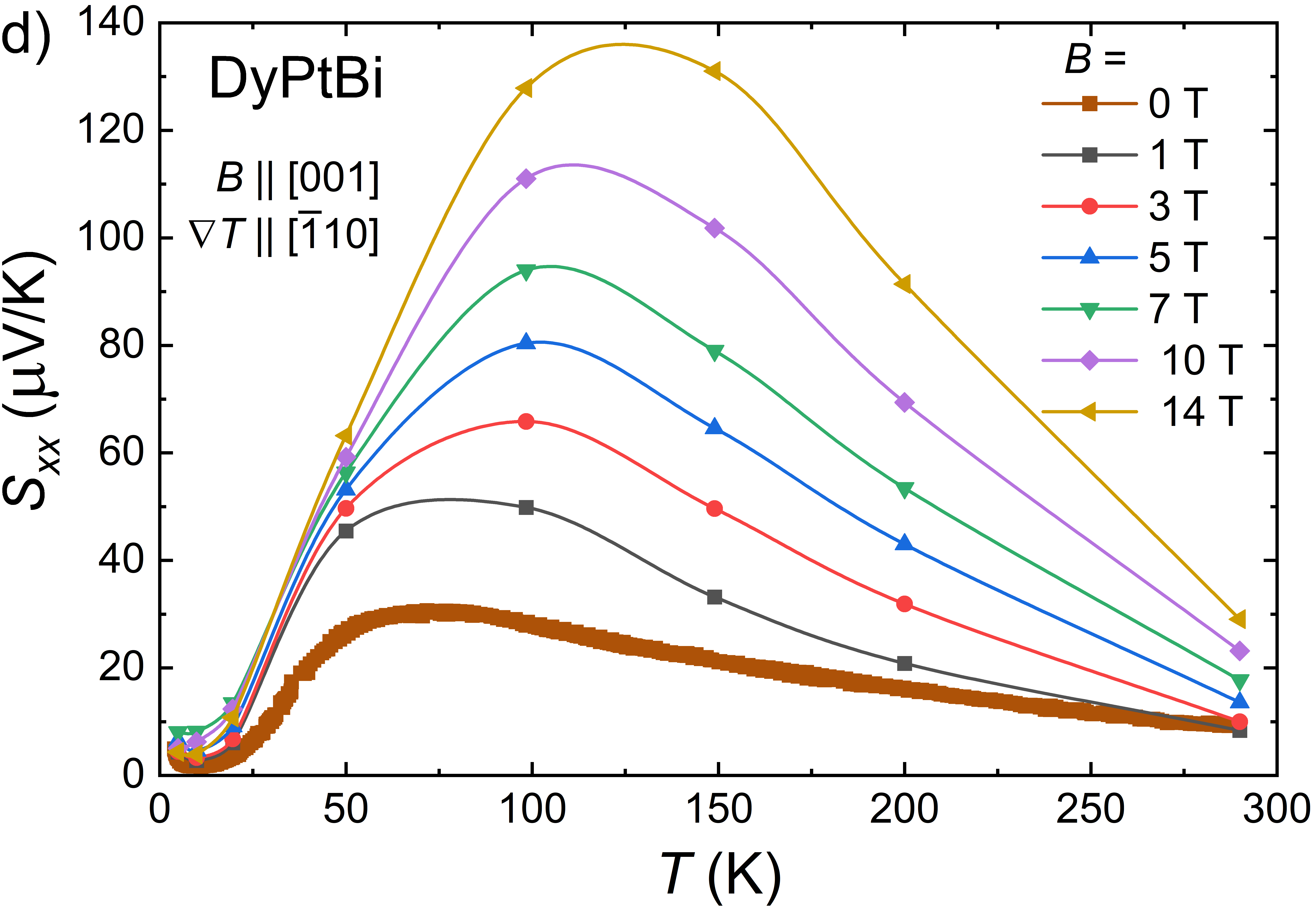}
	\includegraphics[width=0.49\textwidth]{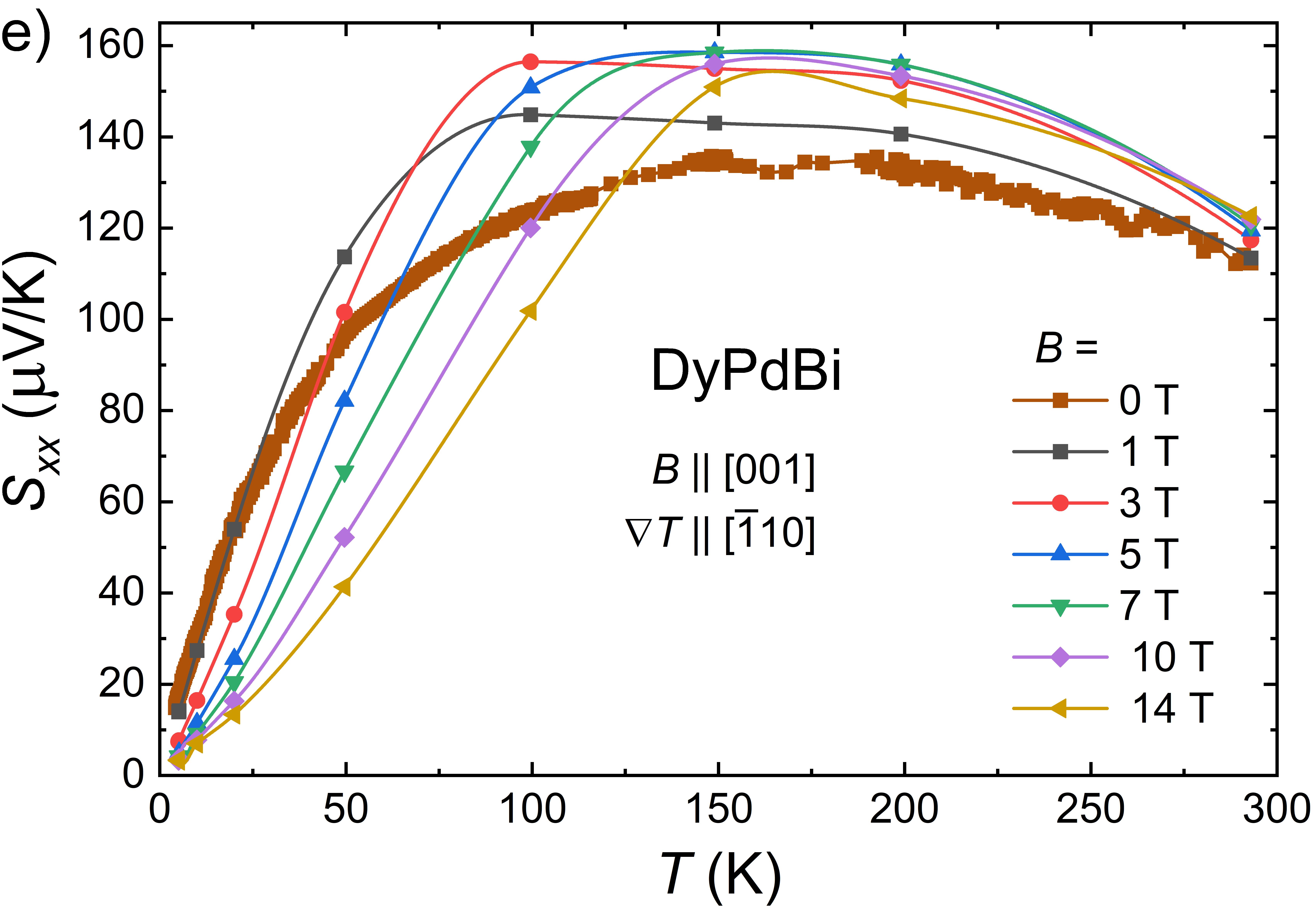}
	\caption{Magnetic field dependence of longitudinal thermopower ($S_{xx}$) of DyPtBi (a, b) and DyPdBi (c) at several different temperatures. $S_{xx}$ as a function of temperature at zero magnetic field and at several different values of applied magnetic field for DyPtBi (d) and DyPdBi (e). Temperature gradient was applied along [$\bar{1}10$] crystallographic direction, magnetic field was applied along [001] crystallographic direction.
		\label{Sxx_vs_B_T}}
\end{figure}

Figure\,\ref{Sxx_vs_B_T}a,b,c depicts the magnetic field ($B$) dependence of $S_{xx}$ of DyPtBi and DyPdBi at various temperatures. 
For both samples the measurement geometry was identical, the temperature gradient ($\nabla T$) and $\textbf{B}$ were applied along [$\bar{1}10$] and [001] crystallographic directions, respectively.
It is evident that the magnetic field has a pronounced impact on $S_{xx}$ of both compounds, although the effect differs between them.  
In general, at $T\geq50$\,K (see Fig.\,\ref{Sxx_vs_B_T}a), $S_{xx}$ of DyPtBi nonmonotonically increases with increasing magnetic field across the entire range of $T$, with the exception of $T=290$\,K and $T=200$\,K at which $S_{xx}$ exhibits a slight decrease in the low magnetic field region. 
At $T=290$\,K this region of magnetic field is larger than at $T=200$\,K.
$S_{xx}(B)$ dependences measured at $T\leq20$\,K (Fig.\,\ref{Sxx_vs_B_T}b) differ notably from those observed at higher $T$ and from those observed for DyPdBi. 
At $T\leq20$\,K in higher magnetic fields, there is a pronounced anomaly in the form of wide peak with the maximum position changing with $T$.  
The anomalies in the similar magnetic field region were also observed in $S_{yx}(B)$, $\rho(B)$ and $\rho_{yx}(B)$, to see this more clearly $\text{d}\rho/{\text{d}}B(B)$ and $\text{d}\rho_{yx}/{\text{d}}B(B)$ plots are shown in the Supporting Information, Fig.\,S2. 
Probably, this anomaly can be attributed to the anomalous Nernst effect (see details below) or to the changes in electronic structure induced by magnetic field. 
Some residual traces of the anomaly are still visible at $T=50$\,K, but at higher $T$, the anomaly disappears and huge increase of $S_{xx}$ with increasing magnetic field is observed. 
For example, the largest value of $S_{xx}$ we observed for DyPtBi is $134\,\upmu\rm{V/K}$ (at $T=149$\,K, in $B=14$\,T), so we observed 540\% growth of $S_{xx}$ with the respect to the value in zero field ($S_{xx}=21\,\upmu\rm{V/K}$). 
This is much larger than the increase we observed for DyPdBi and that reported for TbPtBi.\cite{Wang2022h}
The value $S_{xx}=134\,\upmu\rm{V/K}$ is almost two times smaller than $S_{xx}=251\,\upmu\rm{V/K}$ of TbPtBi but it is comparable to those observed for several topological semimetals, Mg$_3$Bi$_2$,\cite{Feng2022a} Cd$_3$As$_2$\cite{Xiang2020} and TaAs$_2$.\cite{Hu2025}

$S_{xx}$ of DyPdBi behaves in a manner different from that of DyPtBi.  
In the temperature range from 5.1 to at least 199\,K and in low $B$, $S_{xx}$ increases with increasing $B$. 
After achieving a maximum at a particular value of magnetic field ($B_{max}$), $S_{xx}$ starts to decrease with further increase of magnetic field, and tends to saturate at the lowest temperatures. 
For DyPdBi, the highest value of $S_{xx}=159\,\upmu\rm{V/K}$ was observed at $T=149$\,K in $B=6$\,T. 
This represents a mere 18\% increase with respect to the value of $S_{xx}=135\,\upmu\rm{V/K}$ observed at $B=0$\,T at the same temperature. 
For DyPdBi and DyPtBi, $S_{xx}(B)$ curves display a similar pattern to corresponding $MR(B)$ dependences (see Fig.\,\ref{MR_Hall}a and Fig.\,S4 in Supporting Information). 
Quite similar behavior of $S_{xx}(B)$ curves to that demonstrated by DyPdBi was previously observed for TbPtBi\cite{Wang2022h} at low $T$, with the difference that in high $B$, $S_{xx}$ of TbPtBi begins to increase with increasing $B$. 

Apart from the large values of $S_{xx}$ observed in DyPtBi, a particularly salient feature is the absence of saturation of $S_{xx}$ in high magnetic fields. A similar lack of saturation was also observed for DyPdBi, although only at $T\geq10$\,K (Fig.\,\ref{Sxx_vs_B_T}).
According to semiclassical theory, $S_{xx}$ should initially increase in low magnetic fields and then saturate at higher fields,\cite{Liang2013} as exemplified by Cd$_3$As$_2$.\cite{Liang2017}
However, many topologically trivial and non-trivial materials show non-saturating $S_{xx}$ even in high magnetic fields, for example, Bi,\cite{Spathelf2022} WTe$_2$,\cite{Pan2022} NbP,\cite{Scott2023} TaAs$_2$,\cite{Hu2025} Bi$_{88}$Sb$_{12}$\cite{Pan2025} and TbPtBi.\cite{Wang2022h}
Recent investigations have made notable progress in understanding this behavior.\cite{Feng2021,Spathelf2022,Pan2025} 
It should be also noted that none of models for magneto-Seebeck effect discussed in these works could fully describe the experimental data but gave only qualitative descriptions. 
Further research in this area is therefore highly desirable.
Nevertheless, in the case of topologically trivial compensated semimetal bismuth, the unsaturated magneto-Seebeck effect arises from Landau quantization, which changes the electronic structure once the system reaches the extreme quantum limit.\cite{Spathelf2022}
Similarly, it has been concluded that entering the extreme quantum limit by TaAs$_2$ results in a linear magnetic filed dependence of $S_{xx}$.\cite{Hu2025}
In our case, this explanation can be dismissed, as both DyPdBi and DyPtBi do not reach the quantum limit within the accessible range of magnetic fields.
An alternative mechanism responsible for non-saturating $S_{xx}$, involving $B$-induced changes to the electronic structure via Zeeman effect, has been proposed in Bi$_{88}$Sb$_{12}$.\cite{Pan2025}
Given previous reports suggesting a significant role of Zeeman splitting in $RE$PtBi compounds,\cite{Hirschberger2016a} it is reasonable to consider it as a contributing factor to the large and non-saturating $S_{xx}$ observed in both DyPtBi and DyPdBi. 
Another factor may be the semimetallic nature of these compounds.\cite{Feng2021} 
Moreover, the character of $S_{xx}(T)$ curves may suggest that the degree of carrier compensation in DyPtBi is larger than in DyPdBi (see Supporting Information), which is further supported by the results of Hall effect analysis (see Fig.\,S6 in Supporting Information).
Interestingly, DyPdBi also shows a decrease in $S_{xx}$ with increasing $B$, indicating additional mechanism. 
This decrease may arise from the magnetic-field-induced changes in the Fermi surface due to Landau quantization\cite{Spathelf2022} or Zeeman effect\cite{Pan2025} or may come from the reduction of spin-dependent scattering.\cite{Fujishiro2018}  
The latter appears to be the most plausible explanation, particularly in light of the pronounced negative magnetoresistance observed in DyPdBi (see Fig.\,\ref{MR_Hall}), which has been attributed to the suppression of spin scattering.\cite{Pavlosiuk2019}
However, according to the literature reports,\cite{Fujishiro2018,YayuWang2003} this mechanism leads to a smaller suppression of $S_{xx}$ than we observed for DyPdBi. 
Therefore, it is most probably that Zeeman effect\cite{Pan2025} impacts $S_{xx}$ in DyPdBi as well.     

The temperature dependence of $S_{xx}$ is shown in Fig.\ref{Sxx_vs_B_T}d,e for zero and several finite values of applied $B$. 
In DyPtBi, the maximum in $S_{xx}$, which occurs at $T=77$\,K in $B=0$\,T, shifts to the region of higher $T$ with increasing $B$, similar to the behavior observed in Mg$_2$Bi$_3$.\cite{Feng2022a}
However, it is worth noting that the maxima in DyPtBi occur at much higher temperatures (100-150\,K, depending on the value of magnetic filed) than those reported for other well-compensated topological semimetals such as Mg$_2$Bi$_3$\cite{Feng2022a} and NbSb$_2$,\cite{Li2022d} where they are located well below 100\,K. 
This behavior is more comparable to NbP,\cite{Fu2018b,Scott2023} where the position of the maximum has been linked to the magnitude of the Fermi temperature,\cite{Scott2023} or to TbPtBi, for which peak in $S_{xx}(T)$ has been observed just above 200\,K.               
For DyPdBi, $S_{xx}(T)$ in $B$ looks more complex due to the fact of the complex behaviour of $S_{xx}(B)$. 
Increase in magnetic field leads to an increase in $S_{xx}$ when compared to the values of $S_{xx}$ at $B=0$\,T only in the higher temperature region.
The borders of this region also change with magnetic field changes. 
In turn, at low temperatures, $S_{xx}$ in magnetic field decreases when compared to the values of $S_{xx}$ recorded in $B=0$\,T. 

\subsection{Transverse magneto-thermoelectric effect}

\begin{figure}[h]
	\includegraphics[width=0.49\textwidth]{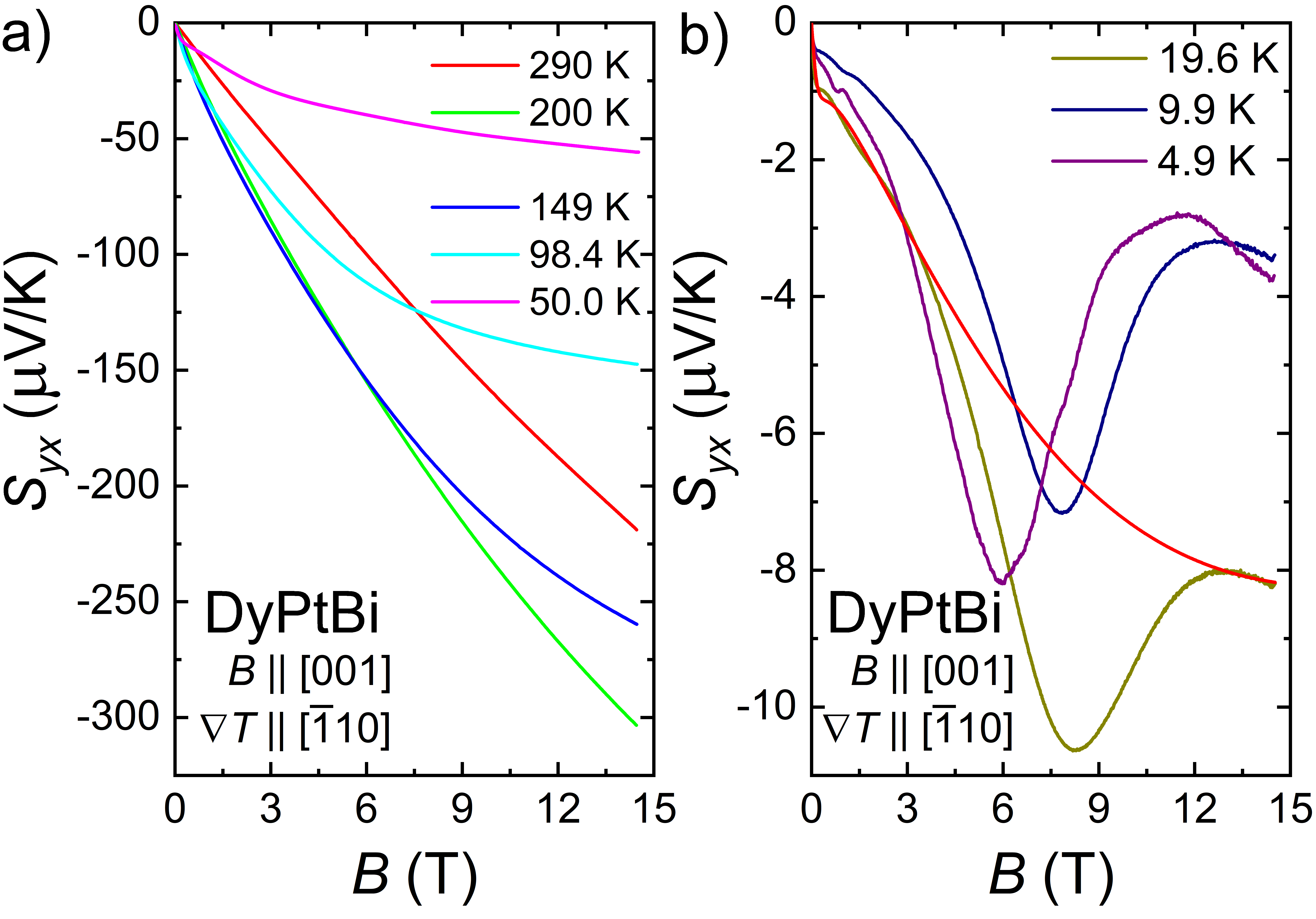}
	\includegraphics[width=0.49\textwidth]{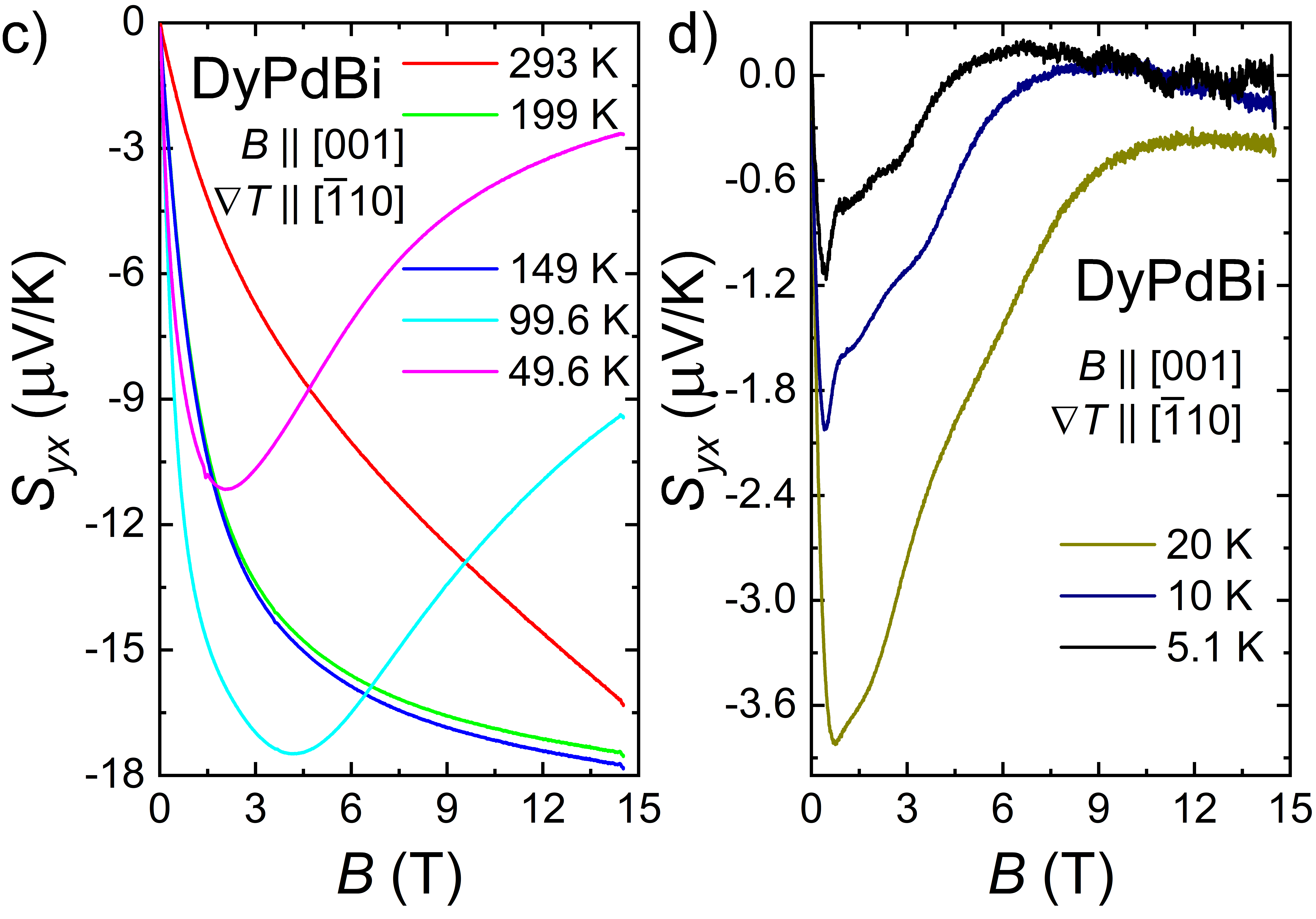}
	\includegraphics[width=0.49\textwidth]{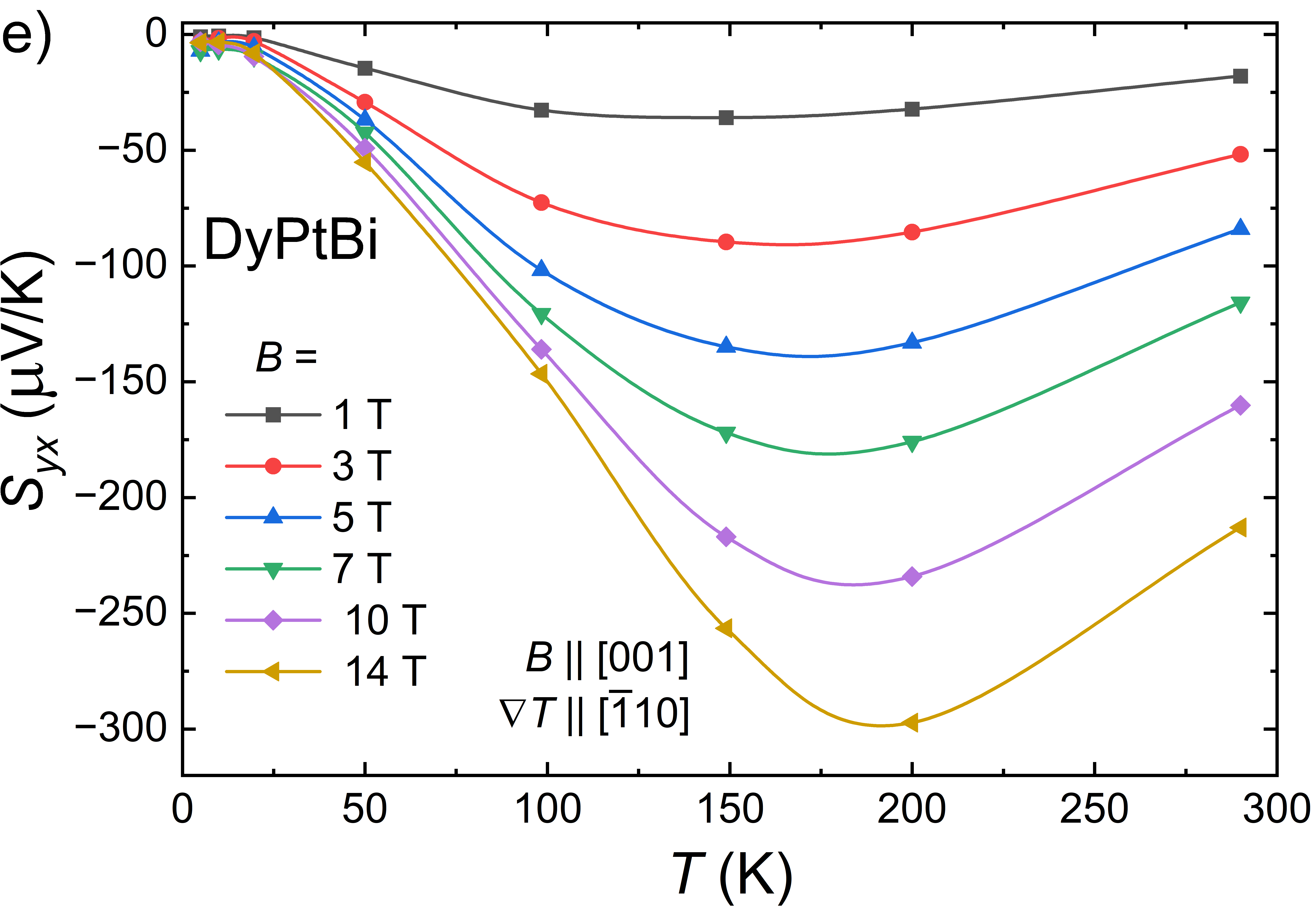}
	\includegraphics[width=0.49\textwidth]{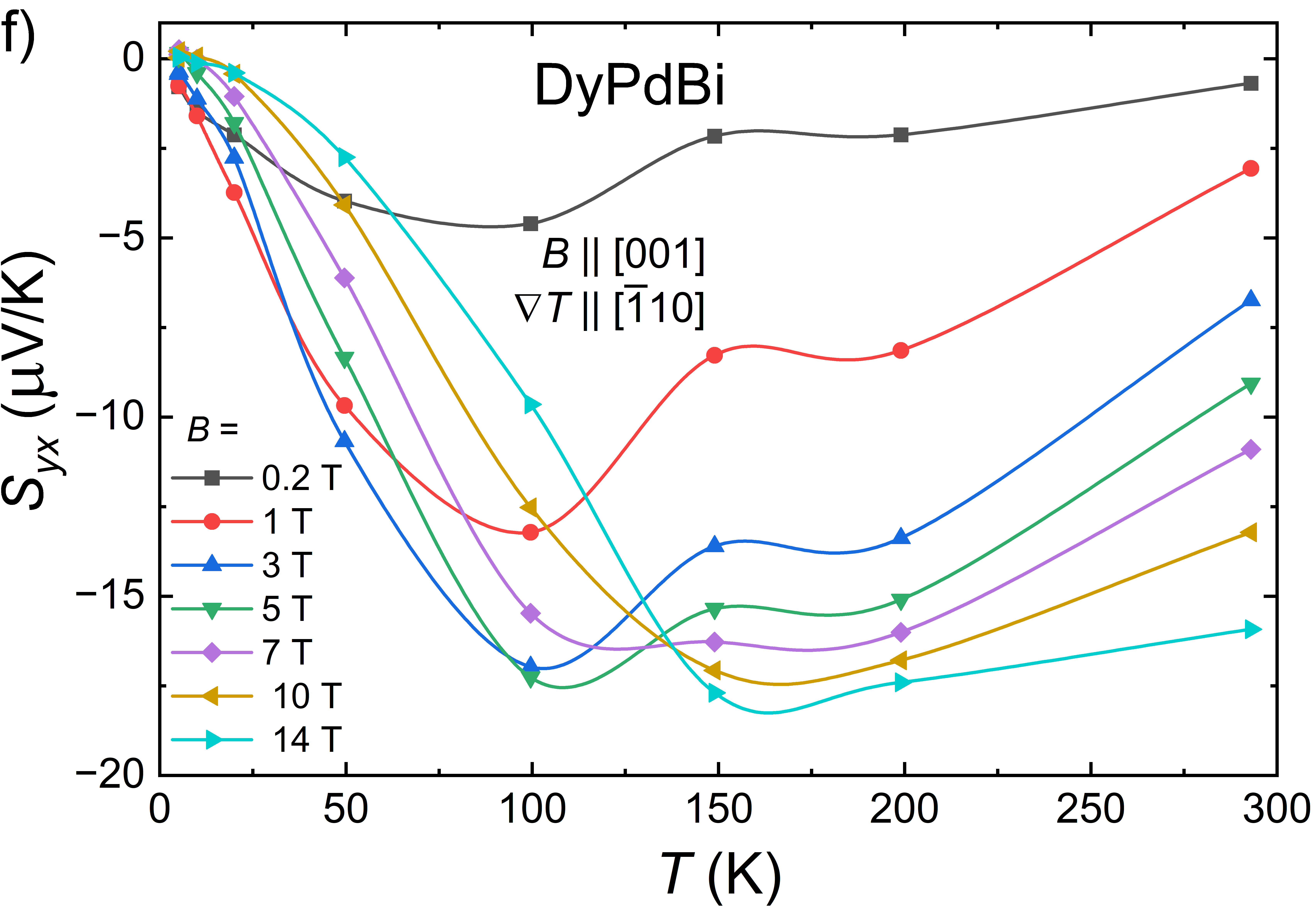}
	\caption{Nernst effect in DyPtBi and DyPdBi. The magnetic field dependence of the transverse thermopower ($S_{yx}$) of DyPtBi (a, b) and DyPdBi (c) at several different temperatures. $S_{yx}$ as a function of temperature at several different values of applied magnetic fields for DyPtBi (e) and DyPdBi (f). The temperature gradient was applied along [$\bar{1}10$] crystallographic direction, and the magnetic field was applied along [001] crystallographic directions. 
		\label{Sxy_vs_B_T}}
\end{figure}

The transverse thermopower due to the Nernst effect ($S_{yx}$) is presented as a function of $B$ in Fig.\ref{Sxy_vs_B_T}a-d for DyPtBi and DyPdBi. 
Similarly to $S_{xx}(B)$, both compounds demonstrate notable differences in the behaviour of $S_{yx}(B)$, which also strongly varies with $T$. 
For DyPtBi, at $T<20$\,K, there is a pronounced anomaly in the $S_{yx}(B)$ curve, the maximum of which varies with temperature (see Fig.\,\ref{Sxy_vs_B_T}b). 
The position of this anomaly is in good agreement with the anomalies in $S_{xx}(B)$, $MR(B)$ and Hall effect data, and we attribute it to the anomalous Nernst effect (ANE) (see Supporting Information). 

At $T\geq20$\,K, in the entire range of magnetic field $|S_{yx}|$ increases with increasing $B$. 
The largest observed absolute value of $S_{yx}$ for DyPtBi is $297\,\upmu\rm{V/K}$ (at $T=200$\,K and in $B=14$\,T), which is 3 times larger than value observed in the same conditions and more than 2 time larger than the maximum of $S_{xx}$ we observed for DyPtBi. 
The observed $|S_{yx}|$ is larger than the maximal value of $S_{yx}=225\upmu\rm{V/K}$ reported for TbPtBi.\cite{Wang2022h}        

At first glance $S_{yx}(B)$ of DyPdBi (see Fig.\ref{Sxy_vs_B_T}c,d) looks more complex compared to DyPtBi (see Fig.\ref{Sxy_vs_B_T}a,b) at $T\geq20$\,K.
At all temperatures, in the low magnetic field region, $|S_{yx}|$ increases with increasing $B$. 
As $T$ increases, this magnetic field region extends to higher magnetic field values and for $T\geq149$\,K $|S_{yx}|$ increases in the entire range of $B$. 
At $T\leq100$\,K, $|S_{yx}|$ achieves a local maximum and subsequently starts to decrease continuously up to the highest $B$.  
These specific shape of measured $S_{yx}(B)$ curves at $T=50$\,K and $T=100$\,K resembles that typical of ordinary Nernst effect (ONE).   
At $T\leq20$\,K, $S_{yx}$ demonstrates a saturation in the highest magnetic fields. 
The temperature evolution of $S_{yx}(B)$ exhibits a similar behavior to that observed in topological crystalline insulator Pb$_{1-x}$Sn$_x$Se in Ref.\cite{Liang2013}
In comparison to DyPtBi, the magnitude of the $|S_{yx}|$ of DyPdBi is considerably smaller. 
For example, the largest observed $S_{yx}$ for DyPdBi was found to be $-17.7\,\upmu\rm{V/K}$ at $T=149$\,K in $B=14$\,T. 
Additionally, at low $B$ and at $T\leq20$\,K, we observed some irregularities in the $S_{yx}(B)$ for DyPdBi, which can be attributed to the anomalous Nernst effect, but their magnitudes are smaller than those we observed for DyPtBi.      

In order to gain a different perspective on the $S_{yx}$ data, we plotted them as a function of $T$ for several constant values of $B$ (see Fig.\,\ref{Sxy_vs_B_T}e,f). 
$S_{yx}(T)$ looks similar to the dependences reported for other semimetals, including NbSb$_2$,\cite{Li2022d} Mg$_3$Bi$_2$\cite{Feng2023} and Mg$_2$Pb.\cite{Chen2021l}
At low $T$, $|S_{yx}|$ increases with increase of $T$, reaching a maximum in the range of 150-200\,K (depending on the magnetic field value) and then decreases with increasing $T$. 
In contrast to the behavior observed in previously studied semimetals,\cite{Li2022d,Feng2023,Chen2021l} DyPtBi demonstrates a maximum in $|S_{yx}|(T)$ at higher temperatures and also shows larger room-temperature values of $|S_{yx}|$ than the above-mentioned semimetals. 
The behavior of $|S_{yx}|(T)$ in DyPdBi is more complex, nevertheless a clear maximum can be observed for each curve measured at a specific value of $B$. 
As the magnetic field increases, the position of this maximum shifts to higher $T$.

We now turn to the origin of the particular $S_{yx}(B)$ dependences observed in DyPdBi and DyPtBi. 
According to the semiclassical Boltzmann transport theory $S_{yx}$ of metal can be described as:\cite{Liang2017,Liang2013}
\begin{equation}
	S_{yx}=S_{yx,0}\frac{\mu\!B}{1+(\mu\!B)^2}.
	\label{S_xy_ONE}
\end{equation} 
Several curves representing this equation are shown in Supporting Information, Fig.\,S3. 
The magnetic field at which $S_{yx}$ reaches its maximum, $B_{max}=1/\mu$, serves as a direct measure of mobility. 
As observed, the shape of $S_{yx}(B)$ curves in DyPdBi measured up to at least 100\,K follows the behavior described by Eq.\,\ref{S_xy_ONE} closely. 
Assuming that mobility becomes sufficiently small such that $B_{max}$ lies beyond the experimental field range, even the data measured at $T>100$\,K may be interpreted in the scope of semiclassical theory of metals. 
This is also consistent with the relatively small values of $S_{yx}$ in DyPdBi, which are much smaller than in DyPtBi.

In an ideally compensated semimetal with equal carrier concentration ($n_h=n_e$) and balanced conductivities of holes and electrons ($\sigma_e=\sigma_h$), the Nernst thermopower in the high-field regime ($\mu\!B\gg1$) is given by:\cite{Chen2021l}
\begin{equation}
	S_{yx}=\frac{S_{e,xy}}{2}+\frac{S_{h,xy}}{2}+\frac{\mu\!B(S_{h,xx}-S_{e,xx})}{2}	
\end{equation} 
This equation highlights that additional ambipolar term is proportional to magnetic field, mobility and difference between longitudinal thermopower of holes and electrons, which have opposite signs. 
Therefore, $S_{yx}$ in well-compensated semimetal can attain large values and not saturate, even in a high magnetic field.
In Ref.\,\cite{Chen2021l} it has also been shown that, even when $\sigma_e/\sigma_h=0.1$ or 10, the unsaturated $S_{yx}$ is possible, but with a reduced magnitude compared to the case of ideally compensated semimetals. 
In contrast, when $\sigma_e\gg\sigma_h$ or $\sigma_e\ll\sigma_h$, $S_{yx}(B)$ becomes qualitatively similar to that of usual metal.\cite{Chen2021l}
Keeping in mind that electronic structure calculations, $\rho(T)$ and $S_{xx}(T)$ support the semimetallic nature of both DyPtBi and DyPdBi, the overall behavior of $S_{yx}(B)$ in DyPtBi and DyPdBi can be understood in the context of charge carrier compensation, which is imperfect in both compounds, but the degree of compensation in DyPtBi is larger than in DyPdBi. 
In addition, the degree of carrier compensation in DyPdBi may be so small that its Nernst effect response resembles that of a simple metal.

It is also worth noting that in topological semimetal NdP, non-saturating longitudinal and transverse magneto-thermoelectric responses have also been attributed to nontrivial band topology alone, without invoking compensation effects of trivial bands.\cite{Scott2023} 
Given that both DyPtBi and DyPdBi belong to the topologically nontrivial $RE$PtBi and $RE$PdBi families, the influence of topological features on their magneto-thermoelectric behavior cannot be excluded.

\subsection{Magnetoresistance and Hall effect}

\begin{figure}[h]
	\includegraphics[width=0.49\textwidth]{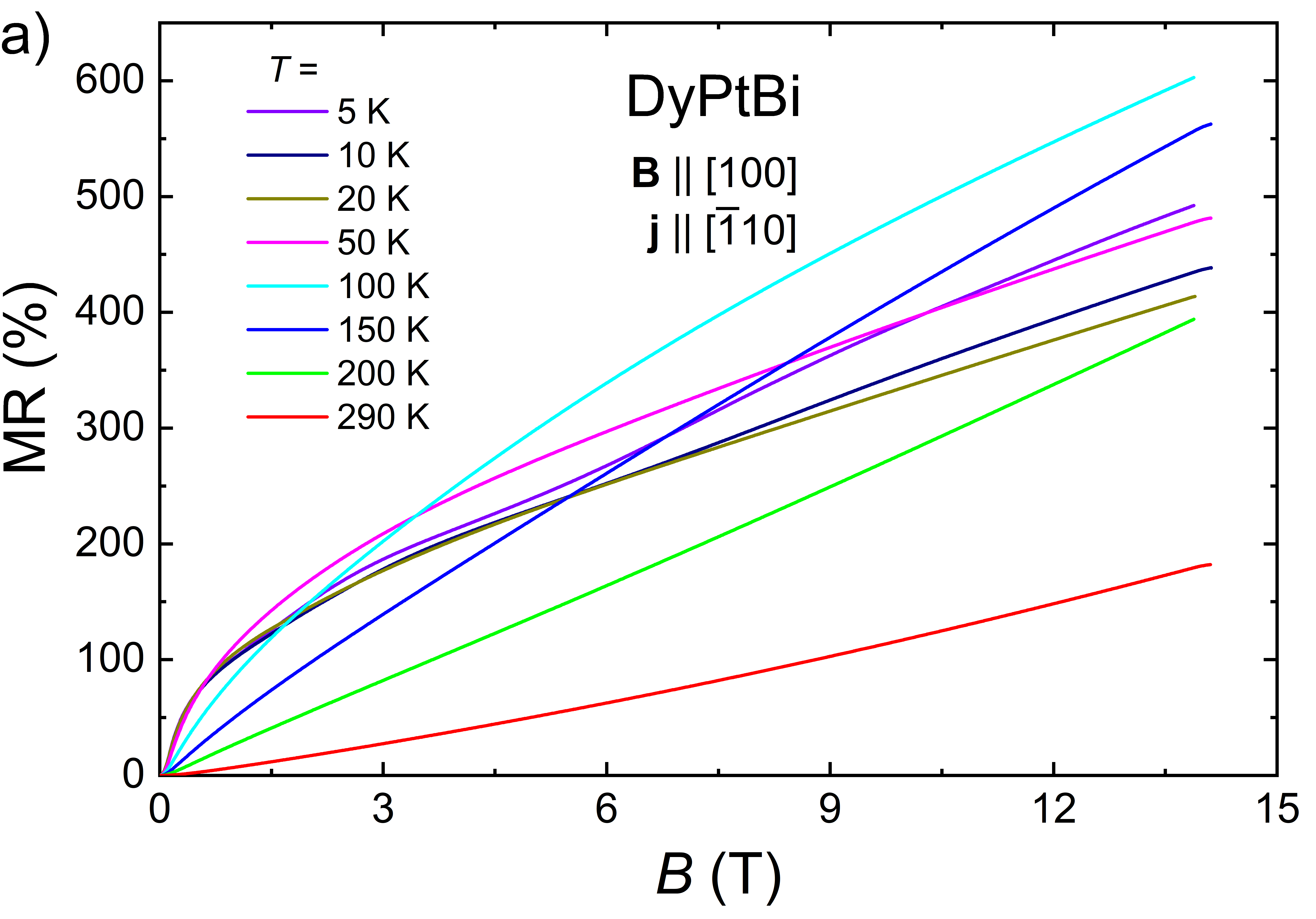}
	\includegraphics[width=0.49\textwidth]{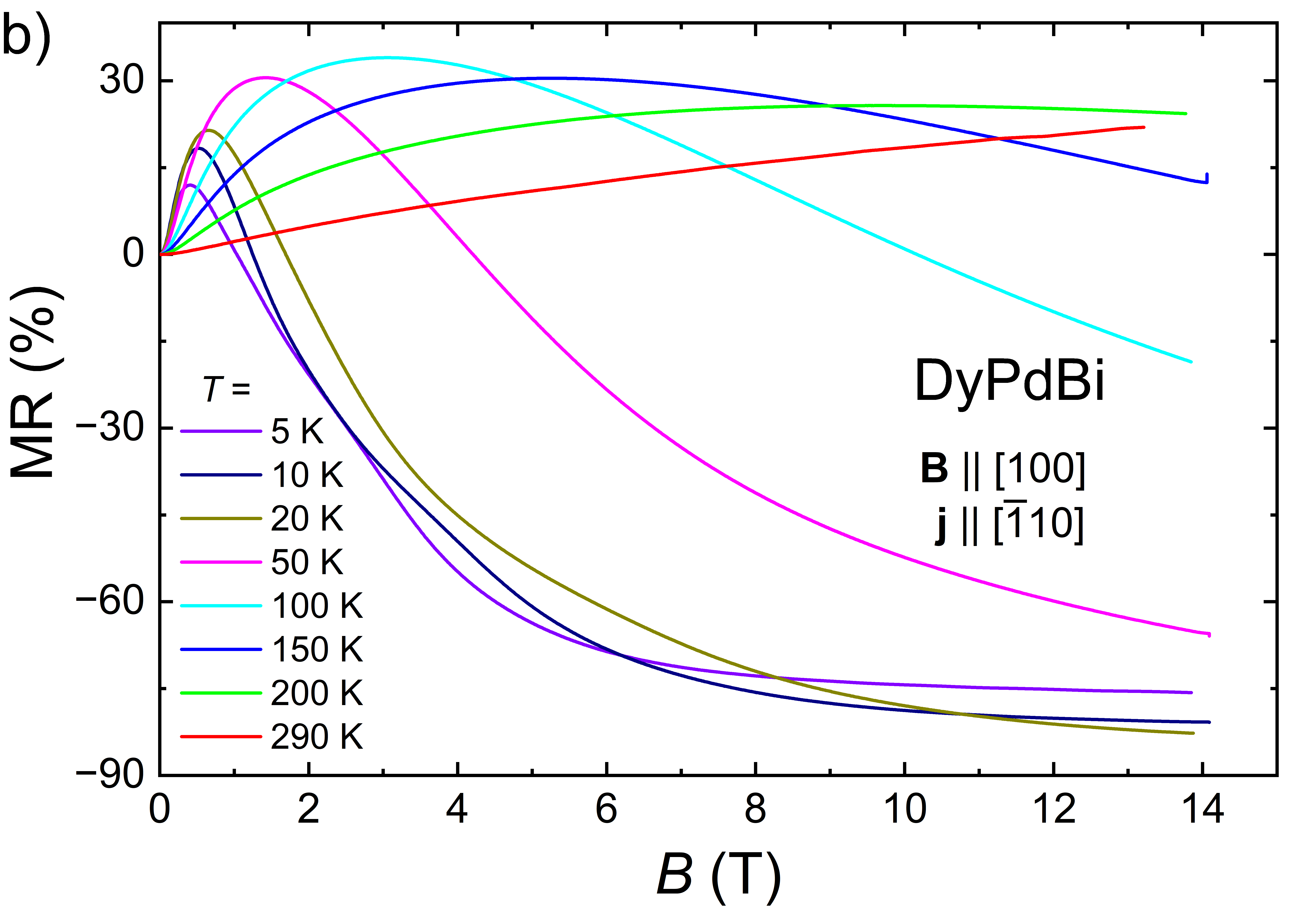}
	\includegraphics[width=0.49\textwidth]{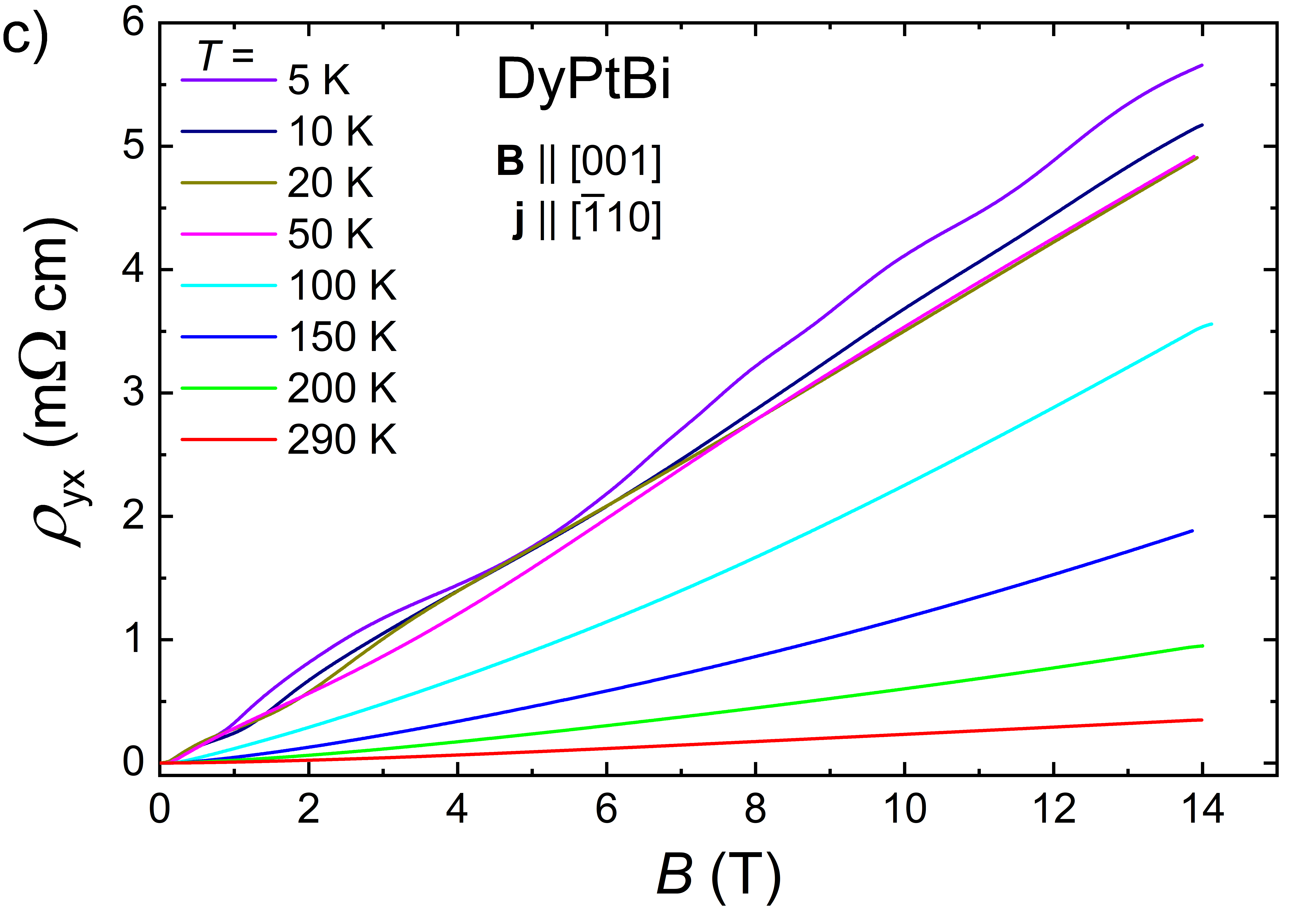}
	\includegraphics[width=0.49\textwidth]{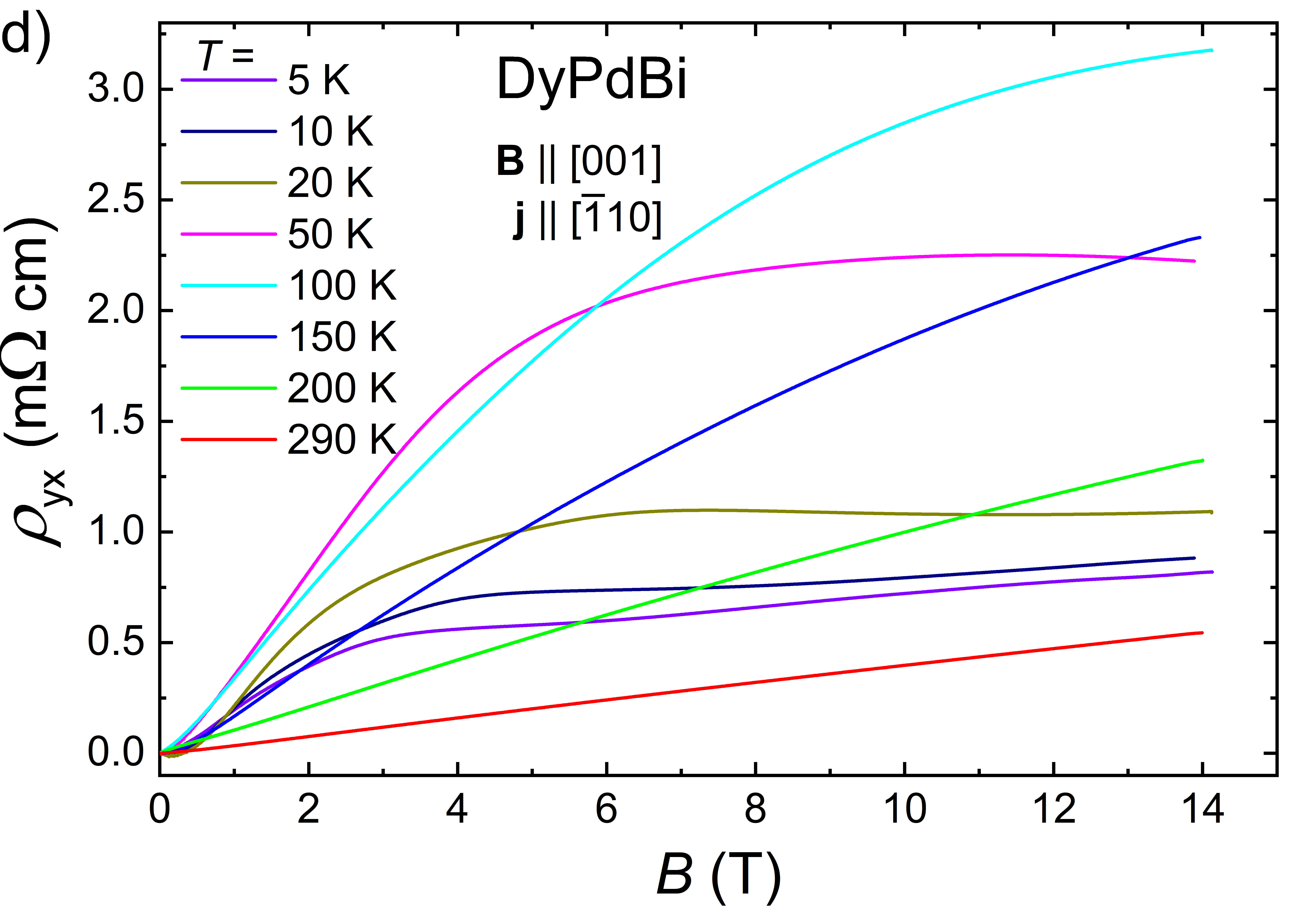}
	\caption{Magnetoresistance isotherms as a function of magnetic field for DyPtBi (a) and DyPdBi (b). Hall resistivity as a function of magnetic field for DyPtBi (c) and DyPdBi (d).
		\label{MR_Hall}}
\end{figure}

In order to estimate the degree of compensation, we studied magnetoresistance (MR) and Hall resistivity ($\rho_{yx}$) as a function of magnetic field at several different temperatures for both compounds (see Fig.\,\ref{MR_Hall}). 
Although, in zero magnetic field $\rho(T)$ of both studied compounds shows rather similar trend (Fig.\,\ref{rho_Sxx}a,b), their MR isotherms display notable differences, only at $T=290$\,K MR($B$) of both compounds look similar, showing positive and non-saturating response in strong $B$ (red lines in Fig.\,\ref{MR_Hall}a,b). 
In turn, DyPdBi shows strong negative MR at $T<100$\,K in high $B$, whereas in low $B$ MR is positive. 
Interestingly, the magnetic field range in which MR remains positive increases and at $T=150$\,K, MR is positive in the entire range of $B$.
The positive slope of MR($B$) for DyPdBi in the entire range of $B$ was observed only for $T=290$\,K (Fig.\,\ref{MR_Hall}b).
In contrast, at lower $T$, DyPtBi demonstrates large and positive MR, which increases with increasing magnetic field in the entire range of $B$ (Fig.\,\ref{MR_Hall}a). 
It should be also noted that MR$(B)$ isotherms of DyPtBi do not resemble those of a compensated semimetal, which would exhibit a convex curvature. 
In DyPtBi, the curvature of MR$(B)$ is concave at $T<200$\,K and only at $T=200$\,K and $290$\,K it becomes convex. 
Therefore, we can speculate that in DyPtBi and DyPdBi, the observed MR reflects the competition between negative (decreasing in $B$) and positive (increasing in $B$) contributions. 
The positive MR contribution most probably originates from the Lorentz force acting on charge carriers. 
In DyPtBi, MR is positive in the entire range of temperatures and magnetic fields. 
In contrast, DyPdBi shows a pronounced negative MR component, particularly at low temperatures and high magnetic fields.   
This difference may be attributed to the varying degrees of compensation in two compounds, as supported by our electronic structure calculations. 
These calculations reveal different electron-hole asymmetry near the $\Gamma$ point and, in case of DyPdBi, the presence of additional electron-like band near the $X$ high-symmetry point close to the Fermi level (see Fig.\,\ref{Electr_str}d,e).  
In turn, the negative MR contribution may be governed by two mechanism: (i) reduction of the spin-disorder scattering\cite{Pavlosiuk2019,Karla1998} and (ii) reconstruction of Fermi surface driven by magnetic field.\cite{Lu2025}

Fig.\,\ref{MR_Hall}c,d shows the results of Hall effect measurements for both compounds. 
In both cases, the hole-type carriers dominate the electron transport, similarly to previous reports.\cite{Pavlosiuk2019,Zhang2020k,Zhu2023b,Zhu2020b,Pavlosiuk2020} 
In the entire temperature range studied, $\rho_{yx}(B)$ demonstrates a non-linear behavior, which may indicate the multi-band mechanism of conductivity. 
However, we were unable to satisfactorily describe $\rho_{yx}(B)$ using the multi-band model due to issues with overparameterization.\cite{Xu2023} 
The simultaneous fitting of both $\rho_{yx}(B)$ and $\rho_{xx}(B)$ also proved unsuccessful, as the pronounced contribution of negative MR is not taken into account by Drude two-band model. 
Only the data measured at $T=290$\,K, could be reliably fitted using two-band model, yielding reasonable parameters. 
These results indicate that DyPtBi shows a higher degree of carrier compensation compared to DyPdBi (see Fig.\,S6 in Supporting Information).

\subsection{Comparison of materials with large longitudinal and transverse thermopowers}

\begin{figure}[h]
	\includegraphics[width=0.95\textwidth]{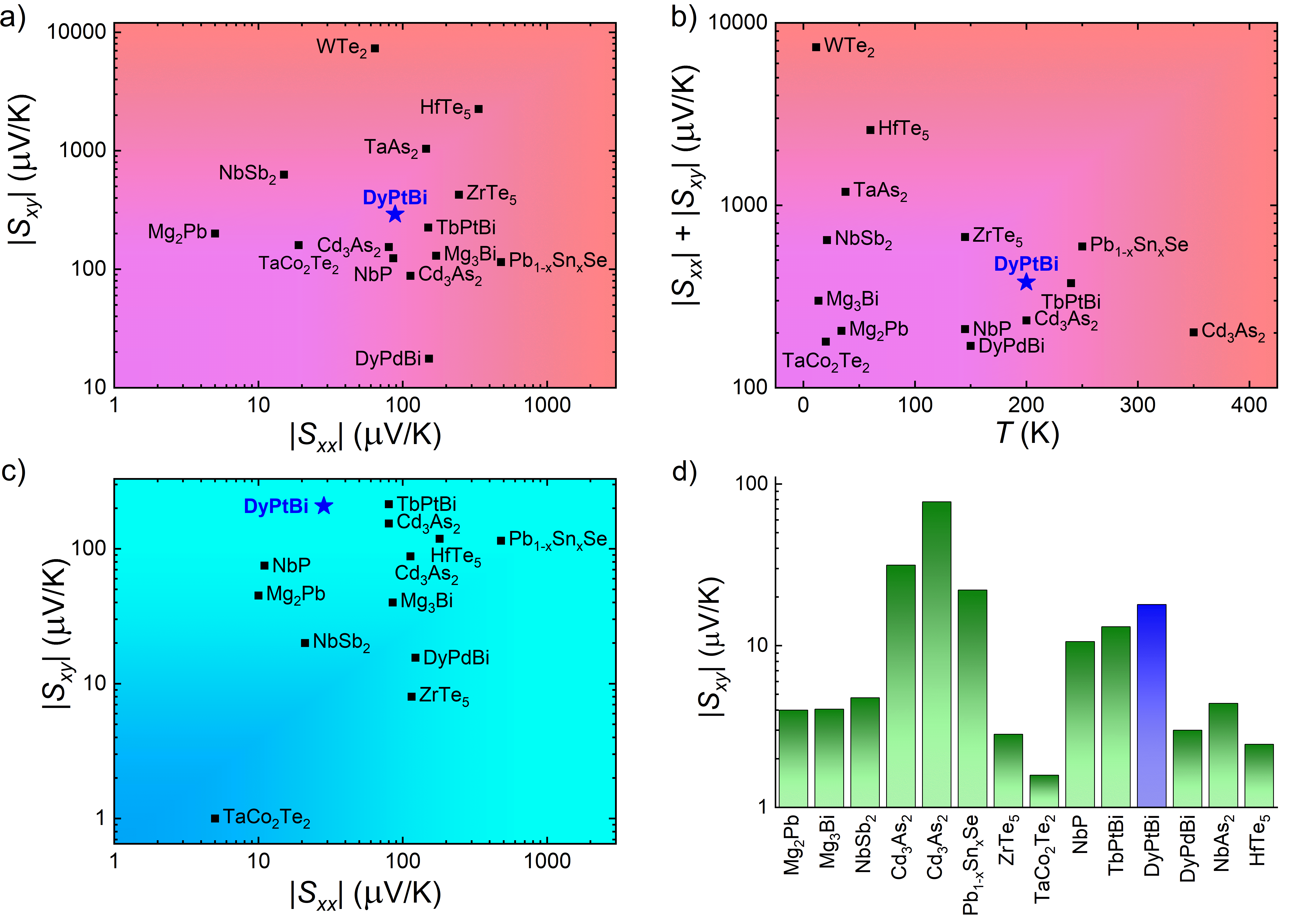}
	\caption{Comparison of the absolute values of the transverse ($S_{yx}$) and longitudinal ($S_{xx}$) magnetothermopower in recently studied topological ambipolar materials. (a) Peak values of $|S_{yx}|$ measured at particular temperatures and magnetic fields, plotted against corresponding $|S_{xx}|$ values obtained under the same conditions. (c) Same as in panel (a), but measurements performed at or near 300\,K. (b) Temperature dependence of $|S_{xx}|+|S_{yx}|$ for the datasets presented in panel (a). (d) $|S_{yx}|$ obtained at $B$=1\,T and at temperatures close to 300\,K. The literature sources of the data presented in this figure are as follows: NbP\cite{Liu2022c}, Mg$_2$Pb,\cite{Chen2021l} NbSb$_2$,\cite{Li2022d} TaCo$_2$Te$_2$,\cite{Gui2024} Cd$_3$As$_2$,\cite{Liang2017,Xiang2020} Mg$_3$Bi,\cite{Feng2022a} ZrTe$_5$,\cite{Zhang2019k} HfTe$_5$,\cite{Jiang2024} TbPtBi,\cite{Wang2022h} WTe$_2$,\cite{Pan2022} TaAs$_2$,\cite{Hu2025} NbAs$_2$\cite{Wu2024} and Pb$_{1-x}$Sn$_x$Se.\cite{Liang2013}       
		\label{Comparison_Sxy_Sxx}}
\end{figure}

In order to benchmark the magneto-thermoelectric effects of DyPdBi and DyPtBi against recently reported materials, we collected literature data and constructed several comparative diagrams, which are presented in Fig.\,\ref{Comparison_Sxy_Sxx}a-d. 
Specifically, we plotted the peak values of $S_{yx}$ (i.e., the maximum values reported for each compound across all measured magnetic field and temperature ranges) as a function of $S_{xx}$ recorded under the same conditions. 
These paired values were taken from the same source for each material to ensure consistency.
As shown in Fig.\,\ref{Comparison_Sxy_Sxx}a, $|S_{yx}|=297\,\upmu\rm{V/K}$ value for DyPtBi is of the same order of magnitude as values reported in several recently studied materials, such as TbPtBi,\cite{Wang2022h} Mg$_2$Pb,\cite{Chen2021l} NbP\cite{} or Mg$_3$Bi$_2$,\cite{Feng2022a} though it is significantly lower than $S_{yx}$ reported for WTe$_2$\cite{Pan2022} and TaAs$_2$\cite{Hu2025}. 
In contrast, DyPdBi shows the smallest $S_{yx}$ among the compared compounds. 
However, both DyPdBi and DyPtBi demonstrate rather large values of $S_{xx}$. 
Taking into account the recent interest in devices that simultaneously may use both $S_{yx}$ and $S_{xx}$ thermopowers,\cite{Scott2023} we calculated the combined value $|S_{yx}|+|S_{xx}|$. 
These results are presented in Fig.\,\ref{Comparison_Sxy_Sxx}b, as a function of temperature at which this sum was achieved. 
Notably, DyPtBi demonstrates one of the largest $|S_{yx}|+|S_{xx}|$ values in the temperature range between 200\,K and 350\,K. 
Only Pb$_{1-x}$Sn$_x$Se\cite{Liang2013} shows a larger value in this range, although it still underperforms compared to materials like WTe$_2$,\cite{Pan2022} TaAs$_2$\cite{Hu2025} and NbSb$_2$,\cite{Li2022d} which reach their maxima below 50\,K. 
Therefore, in Fig.\,\ref{Comparison_Sxy_Sxx}c, we compared $S_{xx}$ and $S_{yx}$ values recorded near ambient condition. 
If data at or near 300\,K were missing, we used the available values, which were measured at the temperature closest to 300\,K reported in the literature. 
DyPtBi stands out for their high values of $|S_{yx}|+|S_{xx}|=235\,\upmu\rm{V/K}$. 
Nevertheless, at $T=290$\,K in DyPtBi $S_{yx}$ is approximately 7 times larger than $S_{xx}$, while the opposite trend is observed in DyPdBi, where $S_{xx}$ is nearly 8 times larger than $S_{yx}$. 
Moreover, the magneto-thermoelectric response of DyPdBi shows much weaker magnetic field dependence compared to DyPtBi (see Figs.\,\ref{Sxx_vs_B_T},\,\ref{Sxy_vs_B_T}). 
Given the strong magnetic-field dependence and large $S_{yx}$ in DyPtBi, we compare it in Fig.\,\ref{Comparison_Sxy_Sxx}d with that of other ambipolar materials recorded near room temperature and in weak magnetic field ($B=1$\,T), which is achievable using permanent magnets. 
Among the materials considered, DyPtBi has one of the largest $|S_{yx}|=18\,\upmu\rm{V/K}$, surpassed only by Pb$_{1-x}$Sn$_x$Se\cite{Liang2013} and Cd$_3$As$_2$.\cite{Liang2017,Xiang2020}
Therefore, these results indicate the potential for chemical tuning in half-Heusler compounds to optimize the compensation degree and achieve large transverse and longitudinal thermopower.

\section{Conclusion}   

Both DyPtBi and DyPdBi investigated in this study retain features of zero-band semiconductors that are particularly susceptible to thermal broadening. 
This thermal broadening leads to imperfect charge carrier compensation, such that both electrons and hole-type carriers contribute to transport phenomena. 
As a result, the observed complex behavior of both $S_{xx}(B)$ and $S_{yx}(B)$ is likely dominated by ambipolar effect. 
However, other mechanisms such as reduction of spin-disorder scattering, Zeeman effect-driven changes in the electronic structure and topologically non-trivial electronic states, may also contribute to magneto-thermoelectric properties.  
This underscores the need for future theoretical research, especially electronic transport and thermoelectric simulations based on Boltzmann transport theory, in order to understand the interplay of these effects more deeply.

Notably, while DyPdBi demonstrates large longitudinal magnetothermopower $S_{xx}=148\,\upmu\rm{V/K}$ and small Nernst effect $S_{yx}=-17\,\upmu\rm{V/K}$, DyPtBi demonstrates large values of both effects ($S_{xx}=92\,\upmu\rm{V/K}$, $S_{yx}=-297\,\upmu\rm{V/K}$) at 200\,K and in $B=14$\,T. 
Our finding highlight the potential of zero-gap semiconductors like $RE$PtBi half-Heusler compounds as a robust platform for studying magnetic-field-induced enhancement of both transverse and longitudinal thermopower even at high temperatures.
Moreover, our results show that appropriate band engineering (through chemical doping) can enhance magneto-thermoelectric responses and shift their peak values closer to room temperature, which opens a new road towards high-performance thermoelectrics capable of operating near room-temperature.  

\section{Experimental Section}
\textbf{Single crystals growth and characterization}:
Single crystals of DyPdBi and DyPtBi were grown using self-flux method (Bi was used as a flux), following the protocols reported in Refs.\cite{Pavlosiuk2019,Pavlosiuk2020} 
The chemical composition of the single crystals was determined to be Dy$_{33.6}$Pt$_{34.25}$Bi$_{32.15}$ and Dy$_{34.6}$Pd$_{32.79}$Bi$_{32.61}$, using a FEI scanning electron microscope equipped with an EDAX Genesis XM4 spectrometer.
The quality of single crystals was checked, and they were oriented by the Laue backscattering technique using a Proto Laue-COS system. 

\textbf{Electronic transport and thermoelectric measurements}:
Electrical resistivity, magnetoresistance and Hall effect measurements were performed on bar-shaped samples using a standard four-probe technique on a Quantum Design PPMS platform. 
Electrical contacts were made of silver wires attached to the sample by silver epoxy paste.

For thermoelectric measurements, the sample was mounted between two phosphor-bronze clamps with Cernox thermometers attached on each side. 
These were used to monitor the thermal gradient produced along the sample by a Micro-Measurements strain-gage heater ($10$\,k$\Omega$ resistance) that was powered by a Keithley\,6221 current source. 
Thermoelectric voltage signals were recorded using an EM Electronics A20a DC sub-nanovolt amplifier connected to a Keithley 2182A nanovoltmeter. 
The temperature dependence of the thermoelectric power was determined using a heater-on/off method, while magnetic field sweeps ($\pm14.5$\,T) were performed with the heater continuously powered.

\textbf{Ab initio calculation}:
The first principles density functional theory (DFT) calculations were performed using the projected augmented-wave (PAW)~\cite{blochl.94} potentials implemented in Vienna Ab initio Simulation Package (VASP)~\cite{kresse.hafner.93,kresse.furthmuller.96,kresse.joubert.99} code.
For the exchange-correlation energy, the generalized gradient approximation (GGA) in the Perdew--Burke--Ernzerhof (PBE) parametrization~\cite{perdew.burke.96} was used.
The energy cutoff for the plane-wave expansion was set to $400$\,eV. 
In the calculations, the experimental values of lattice constant were used: $a=6.6455\,\text{\AA}$ for DyPtBi and $a=6.624\,\text{\AA}$ for DyPdBi.
Dy $4f$-orbitals were treated as a core (valence) states in the case of non-magnetic (magnetic) electronic structure calculations.

\medskip
\textbf{Acknowledgements}
This research was financially supported by the National Science Centre of Poland under projects no. 2021/40/Q/ST5/00066 (D.K.) and 2023/51/D/ST3/01564 (O.P.).

\section*{Author Contributions}

Orest Pavlosiuk: conceptualization, data curation, formal analysis, investigation, methodology, visualization, funding acquisition, writing – original draft, and writing – review \& editing. 
Marcin Matusiak: investigation, data curation, and writing – review \& editing. 
Andrzej Ptok: data curation, formal analysis, investigation, visualization, and writing – review \& editing.
Piotr Wiśniewski: writing – review \& editing. 
Dariusz Kaczorowski: funding acquisition, and writing – review \& editing.



\newpage

\pagenumbering{arabic}
\setcounter{page}{1}
\renewcommand{\thepage}{S\arabic{page}}

\setcounter{figure}{0}
\setcounter{table}{0}
\setcounter{equation}{0}
\renewcommand{\thetable}{S\arabic{table}}
\hyphenpenalty=10000
\exhyphenpenalty=10000
\renewcommand*{\familydefault}{\sfdefault}
\bibliographystyle{naturemag}
\renewcommand{\thefigure}{S\arabic{figure}}
\renewcommand{\theequation}{S\arabic{equation}}

\begin{large}{\bf Supporting Information for} \end{large}\\\\
\begin{Large}{\bf Transverse and Longitudinal Magnetothermopower Promoted by Ambipolar Effect in Half-Heusler Topological Materials}\end{Large}\\\\
\begin{large}Orest Pavlosiuk$^{1,*}$, Marcin Matusiak$^{1,2}$, Andrzej Ptok$^3$, Piotr Wiśniewski$^1$, Dariusz Kaczorowski$^{1,*}$\end{large}\\\\
{\it$^1$~Institute of Low Temperature and Structure Research, Polish Academy of Sciences, ul.~Ok{\'{o}}lna~2, 50-422 Wroc{\l}aw, Poland\\
$^2$~International Research Centre MagTop, Institute of Physics, Polish Academy of Sciences, Aleja Lotnikow 32/46, PL-02668 Warsaw, Poland\\
$^3$~Institute of Nuclear Physics, Polish Academy of Sciences, W. E. Radzikowskiego 152, PL-31342 Krak\'{o}w, Poland}\\

	\subsection*{Electronic structure}

	\begin{figure}[h]
		\includegraphics[width=0.98\textwidth]{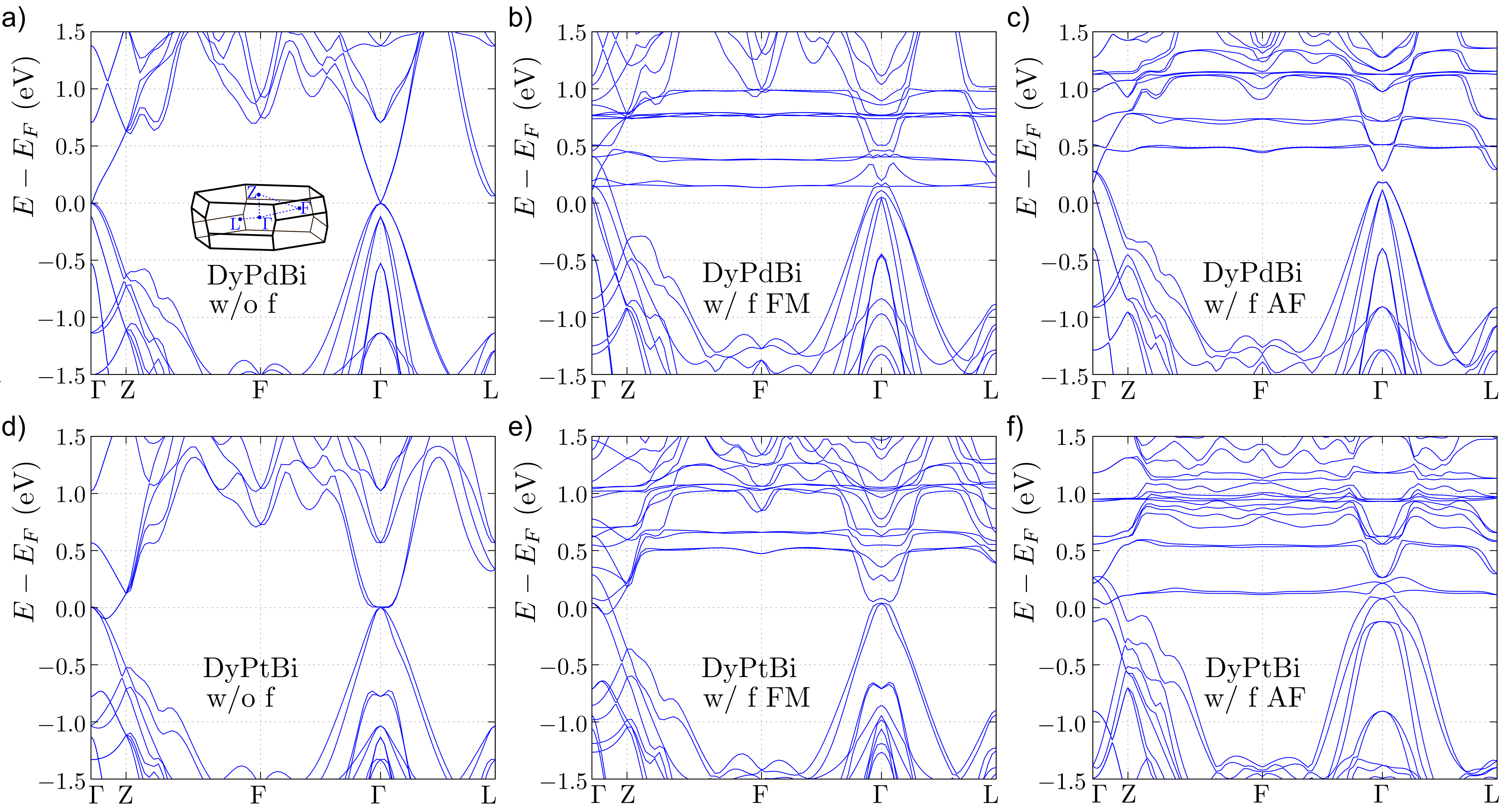}
		\caption{Electronic band structure of DyPdBi (a-c) and DyPtBi (d-f). Panels (a, d)
			present results with Dy $f$ states treated as core states, while panels (b, c, e, f) show results with $f$ states treated as valence states. Ferromagnetic (FM) and antiferromagnetic (AFM) magnetic orderings are assumed in panels (b, e) and (c, f), respectively. Band structures are plotted along high-symmetry directions of the Brillouin zone presented in inset to panel (a). 
			\label{electr_str}}
	\end{figure}

	Although the magneto-transport and magneto-thermoelectric properties were investigated at temperatures at which both studied materials are in the paramagnetic state, magnetic ordering was also taken into account in the calculations (see Fig.\,\ref{electr_str}). 
	As confirmed by neutron diffraction experiments, both compounds show type-II antiferromagnetic structure with a propagation vector of ${\bm{k}}=(1/2, 1/2, 1/2)$.\cite{Pavlosiuk2018,Zhang2020k1}
	The electronic structure for type-II antiferromagnetic order is shown in Fig.\,\ref{electr_str}a and Fig.\,\ref{electr_str}d for DyPdBi and DyPtBi, respectively. 
	For this configuration, the electronic structure of DyPtBi becomes semimetallic-like, the Fermi level intersects the bands of both hole- and electron-type along $\mathrm{\Gamma\!-\!Z}$ line. 
	The results of electronic structure calculations, which are shown in Fig.\,\ref{electr_str}a,\,d were obtained by treating $f\!-\!$electrons as core electrons. 
	However, we also performed calculations including $f\!-\!$electrons explicitly for both antiferromagnetic and ferromagnetic states, the corresponding results are shown in Fig.\,\ref{electr_str}b,\,c,\,e,\,f. 
	As can be seen, the inclusion of $f\!-\!$electrons leads to increase in the density of states around the Fermi level, resulting in electronic structure characteristic of metals.    
	
	\subsection*{Temperature dependences of electrical resistivity and longitudinal thermopower}
	
	For both studied compounds, the broad hump in $\rho(T)$ occurs at almost the same temperature range, indicating that in zero magnetic field both samples are very similar electronically. 
	Previous studies of different samples of another half-Heusler compound GdPtBi have shown that the position of the hump on the temperature scale correlates with the carrier concentration of the particular sample.\cite{Hirschberger2016a1}
	Interestingly, in the narrow-gap semiconductor ZrTe$_5$, which demonstrates quite similar shape of $\rho(T)$ to that observed in our material, the position and amplitude of the resistivity hump are sensitive to the degree of carrier compensation.\cite{Pi2024a} 
	This has been supported by theoretical simulations based on the semiclassical Boltzmann transport framework.   
	
	The overall shape of $S_{xx}(T)$ for DyPtBi (Fig.\,2b) is very similar to that observed previously for half-Heusler phases SmPtBi\cite{Kim2001_1} and DyPtBi,\cite{Mun2016a_1} as well as in a promising thermoelectric material, TaCo$_2$Te$_2$.\cite{Gui2024_1}   
	At room temperature, in DyPtBi, $S_{xx}=9\,\upmu\rm{V/K}$, which is almost identical to the value reported for SmPtBi\cite{Kim2001_1} and is approximately two times smaller than the value reported for DyPtBi by another research group.\cite{Mun2016a_1} 
	With a further decrease in temperature, the value of $S$ increases rapidly, reaching a maximum value of $S_{xx}=31\,\upmu\rm{V/K}$ at $T=77$\,K. 
	This is followed by a rapid decrease of $S_{xx}$ until $T=10$\,K, after which a small increase of $S$ is observed. 
	The origin of the upturn at low $T$ can be attributed to the ambipolar conduction as well.\cite{Shahi2018_1} 
	
	At room temperature, for DyPdBi $S_{xx}=112\,\upmu\rm{V/K}$. 
	As $T$ increases, there is a slight increase in $S$, which reaches its maximum value of $S_{xx}=135\,\upmu\rm{V/K}$ at $T=175$\,K. 
	With a further $T$ decrease, $S_{xx}$ decreases slightly and reaches the same value observed at room temperature at $T=75$\,K. 
	Subsequently, there is a rapid decline in $S_{xx}$ down to $S_{xx}=15\,\upmu\rm{V/K}$ at $T=4$\,K.
	To the best of our knowledge, there is no report concerning the thermoelectric properties studies of $RE$PdBi in single-crystalline form. 
	However, previously, polycrystalline samples of several $RE$PdBi (including DyPdBi) have been studied.\cite{Gofryk2011_1,Mukhopadhyay2019_1}         
	The overall $S_{xx}(T)$ behaviour of our material is generally comparable to that observed for a polycrystalline sample of DyPdBi in Ref.\,\onlinecite{Gofryk2011_1}, with the exception that the maximum on $S(T)$ is observed at lower $T$ if compared to the literature report.\cite{Gofryk2011_1} 
	Additionally, the magnitude of $S_{xx}$ is larger than that of the reported polycrystalline material. 
	This is due to the fact that the mean free path and carrier mobility in single crystalline materials are larger than those observed in polycrystalline materials.  
	
	The substantial difference in the values of $S_{xx}$ for both compounds may be attributed to the different degree of carrier compensation ($n_e/n_h$, where $n_e$ and $n_h$ correspond for charge carrier concentration of electrons and holes, respectively) and/or the ratio of carrier mobility of electrons and holes, $\mu_e/\mu_h$. 
	In DyPtBi, both ratios can be closer to 1 than in DyPdBi, meaning that the contributions of hole- and electron-type carriers to $S_{xx}$ will cancel each other to a greater extent than in the former compound. 
	In general, in DyPdBi the overall behavior of $S_{xx}(T)$ resembles that of pure bismuth, reported and completely explained in the context of its semimetallic nature in Ref.\,\onlinecite{Spathelf2022_1}. 
	For bismuth, it has also been concluded that difference between mobilities of different bands, as well as temperature dependence of electronic structure, plays a crucial role in the understanding of $S_{xx}$ in zero magnetic field.\cite{Spathelf2022_1} 
	
	The high values of the Seebeck coefficient observed for DyPdBi also result in a considerable power factor (${\rm{PF}}=S_{xx}^2/\rho$), for example ${\rm{PF}}=15\,\upmu\rm{W/(K^2\,cm)}$ at room temperature. 
	This value is markedly higher than that observed for DyPtBi, where  ${\rm{PF}}=0.06\,\upmu\rm{W/(K^2\,cm)}$. 
	The obtained value of PF for DyPdBi is comparable to the values reported for the poly-crystalline $RE$PdBi\cite{Gofryk2011_1} and $RE$NiSb\cite{Ciesielski2020} materials.
	
	\subsection*{Low-temperature anomalies in magneto-transport and magneto-thermopower}
	\begin{figure}[h]
		\includegraphics[width=0.75\textwidth]{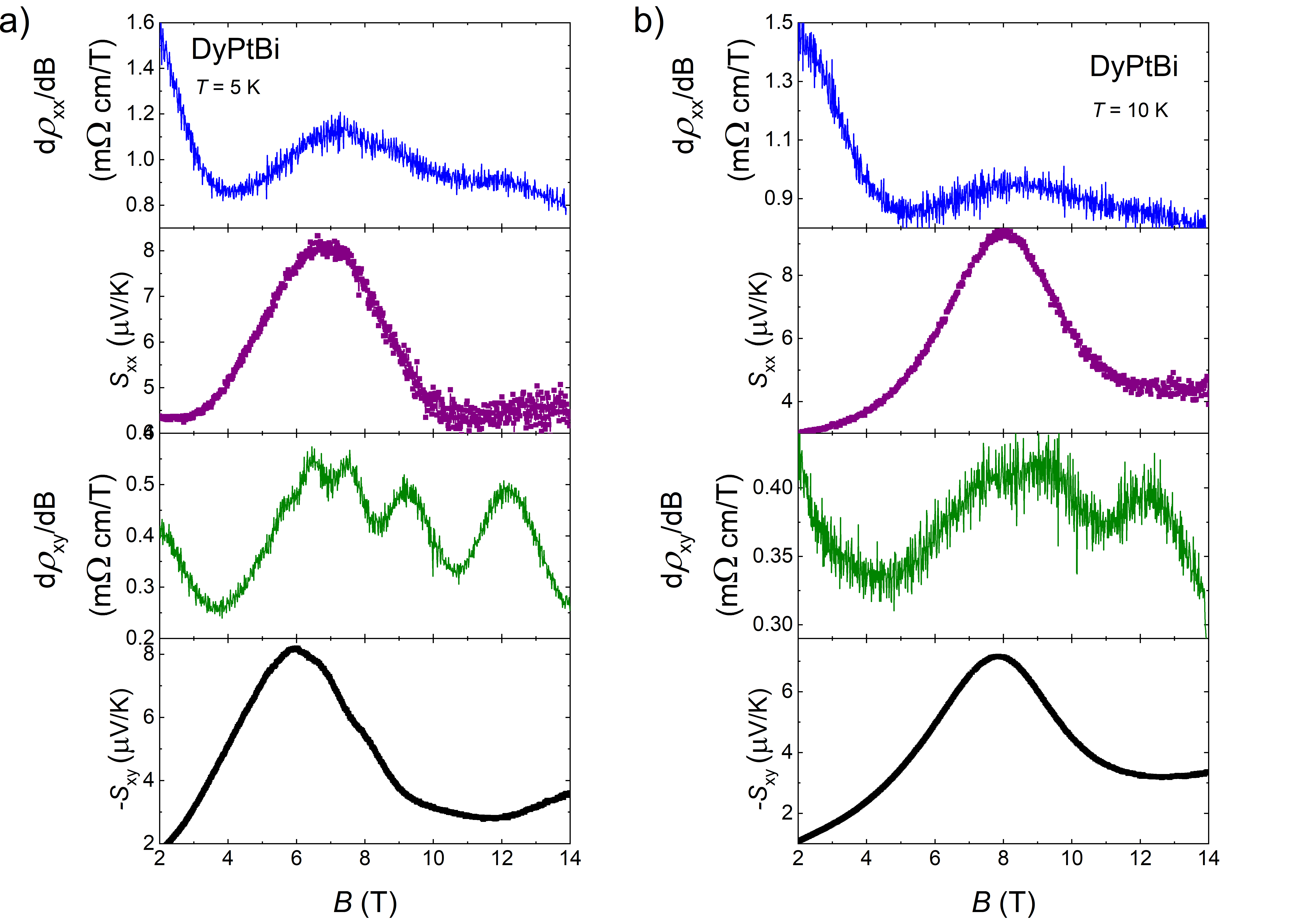}
		\includegraphics[width=0.75\textwidth]{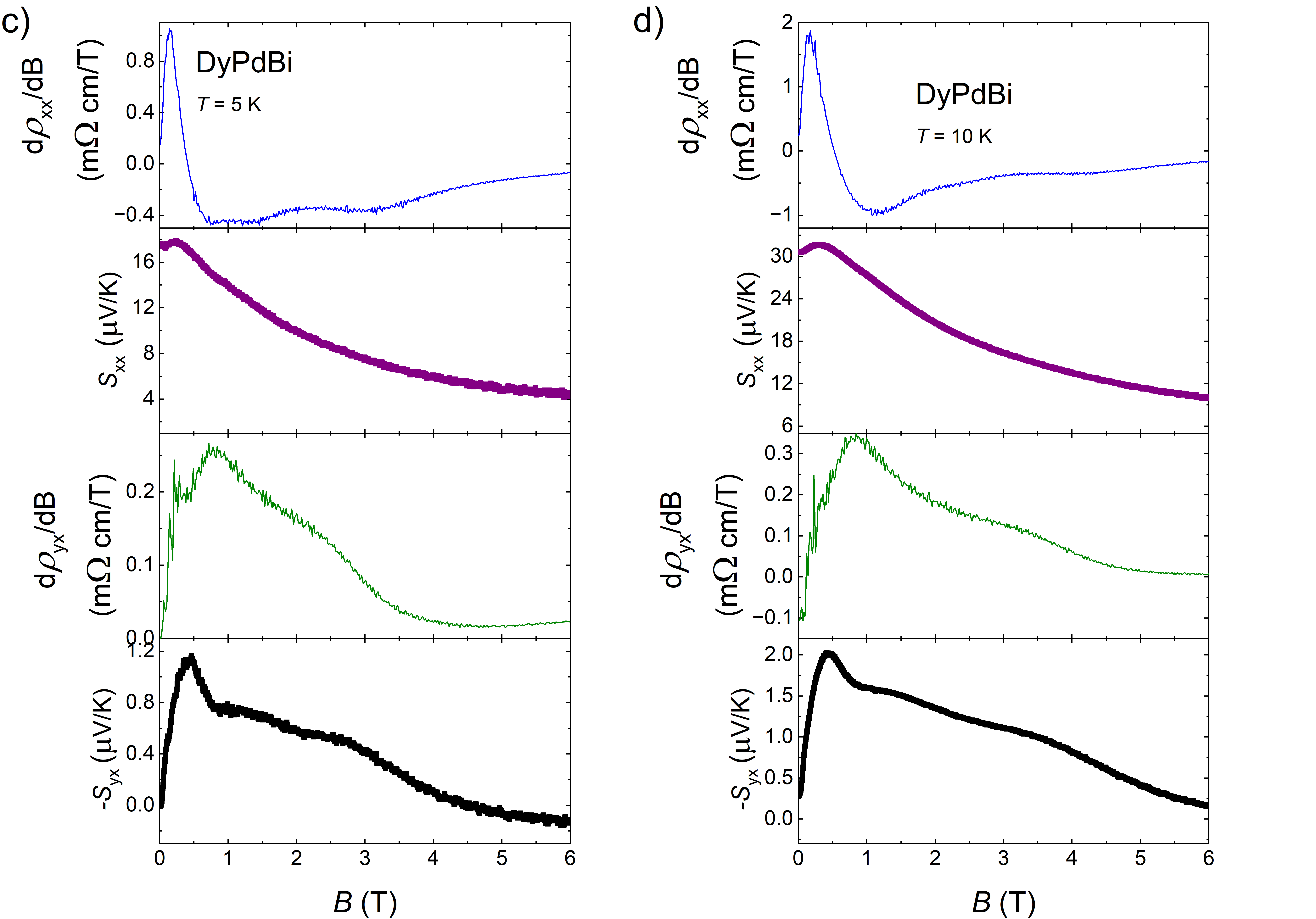}
		\caption{Derivative of electrical resistivity over magnetic field, thermopower, derivative of Hall resistivity over magnetic field and Nernst thermopower as functions of magnetic field for DyPtBi (a, b) and DyPdBi (c, d) at $T=5$\,K (a, c) and $T=10$\,K (b, d). 
			\label{AHE_ANE}}
	\end{figure}

	At $T<20$\,K, the total Nernst thermopower of DyPtBi equals to: $S_{yx}=S_{yx}^O+S_{yx}^A$, where $S_{yx}^O$ represents the ordinary Nernst effect and $S_{yx}^A$ denotes the contribution of ANE. 
	However, the shape of $S_{yx}(B)$ curves indicates that the ANE component in DyPtBi cannot be described by a simple $\approx\!\tanh(B/B_0)$ dependence, where $B_0$ stands for the saturation of ANE, as in case of Cd$_3$As$_2$.\cite{Liang2017_1} 
	Instead, ANE in DyPtBi may show a more complex magnetic field dependence, reminiscent of the behavior observed  in EuCd$_2$As$_2$\cite{Xu2021}, especially considering the complex structure of anomalous Hall effect reported in other $RE$PtBi compounds.\cite{Suzuki2016_1,Pavlosiuk2025_1,Zhu2023b1}
	
	Interestingly, the anomalous contribution is more pronounced in the results of thermoelectric properties measurements than in the data of resistivity tensors measurements. 
	This is due to the fact that, in accordance with the Mott relation, the components of the thermoelectric tensors are proportional to the derivatives of the conductivities tensors over the temperature.\cite{Liang2017_1,Ziman1960} 
	
	\subsection*{Origin of Nernst effect}

	\begin{figure}[h]
		\includegraphics[width=0.49\textwidth]{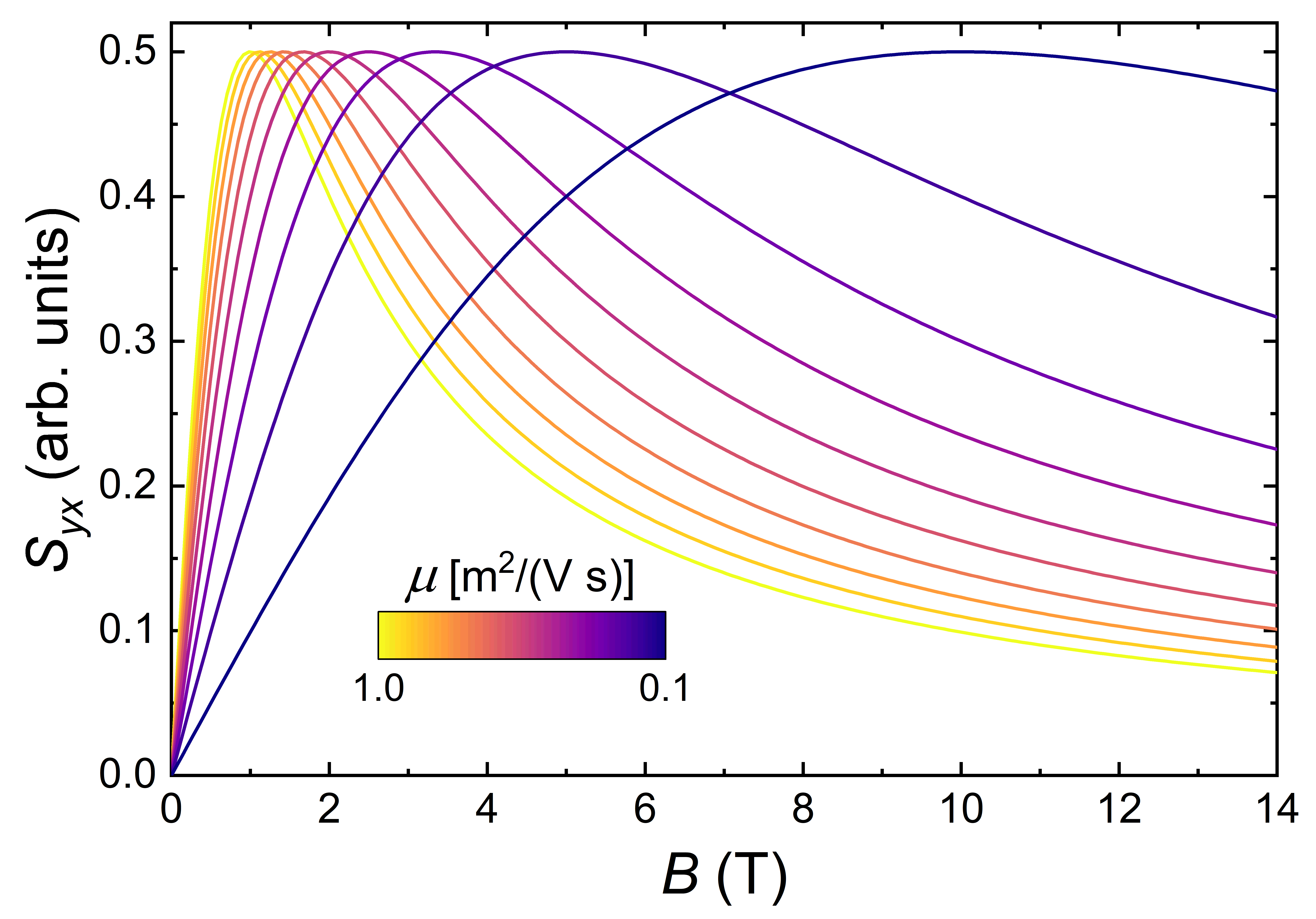}
		\caption{Magnetic field dependence of Nernst thermopower calculated using Eg.\,1 (main text) for several different mobility values. As mobility decreases, the peak position shifts toward higher magnetic fields.
			\label{S_xy_sim}}
	\end{figure}

	In the low magnetic filed regime ($\mu\!B\ll1$) and low temperature limit, Nernst thermopower of a metal can be approximated, based on the semiclassical theory, as:\cite{Behnia2016} $S_{yx}\sim\frac{\pi^2}{3}\frac{k_B^2T}{eE_F}\mu\!B$. 
	In the case of 3D electron gas, $E_F=\frac{\hbar^2}{2m^*}(3\pi^2n)^{2/3}$, implying $S_{yx}\propto\frac{\mu}{n^{2/3}}$. 
	This indicates that Nernst effect will be large in materials with high carrier mobility and small carrier concentration. 
	In the high magnetic field regime ($\mu\!B\gg1$), where electrical transport is governed primarily by cyclotron motion rather than scattering, Nernst thermopower follows $S_{yx}\propto\frac{1}{B}$. 
	Equation 1 in the main text is commonly used to describe the behavior of $S_{yx}(B)$ over the entire range of magnetic fields, and Fig.\,\ref{S_xy_sim} presents $S_{yx}(B)$ curves calculated from this equation for several different values of $\mu$.    
	
	The overall behavior of $S_{yx}(B)$ in DyPtBi can be attributed to the ambipolar effect.\cite{Delves1965_1}
	In the case of perfectly compensated semimetal with identical carriers mobilities of different bands, the Nernst thermopower will be entirely dominated by the ambipolar effect, and also in these conditions, $S_{xx}$ will be zero. 
	However, if compensation is imperfect and/or the carrier mobilities of different bands differ greatly, both $S_{yx}$ and $S_{xx}$ can achieve non-zero values or even reach rather large values.\cite{Chen2021l_1,Markov2019,Feng2021_1} 
	For example, in TbPtBi\cite{Wang2022h1}, both $S_{yx}$ and $S_{xx}$ have similar order of magnitude, with $S_{xx}$ being slightly larger than $S_{yx}$. 
	In turn, in one of the most highly compensated semimetals with ultra-high mobility currently known, WTe$_2$,\cite{Pan2022a} $S_{yx}$ is two orders of magnitude larger than $S_{xx}$. 
	DyPtBi occupies an intermediate position between the two previously discussed materials, but is more closely aligned with TbPtBi.
	
	It has recently been demonstrated that in the well-compensated semimetal Mg$_2$Pb, the overall $S_{yx}(B)$ can be decomposed into two contributions: the first coming from ambipolar effect, which is a linear function of magnetic field, and the second from particular bands (ordinary Nernst effect, ONE), which shows more complex behaviour, similar to that described by Eq.\,(1) in main text.\cite{Chen2021l_1}
	The ambipolar effect was observed to have a greater impact on the total $S_{yx}(B)$ in Mg$_2$Pb relative to the latter contribution. 
	DyPtBi and DyPdBi have a smaller degree of compensation than Mg$_2$Pb and the relative contributions of the ambipolar effect and the particular bands will therefore differ.
	
	In Ref.\,\onlinecite{Feng2021_1}, it has been shown that below the extreme quantum limit in both topologically trivial and nontrivial but well compensated semimetals ($n_e/n_h=0.99$), $S_{xx}(B)$ follows a similar trend: it remains $B$-independent at low fields ($B<B_1$), increases quadratically in intermediate fields, and eventually saturates at high fields ($B>B_H$). 
	The characteristic magnetic field scales, $B_1=m/(e\tau)$ and $B_H\sim\!B_1\times\!n_e/\Delta\!n$, depend on the effective mass ($m$), scattering time ($\tau$) and degree of compensation $\Delta\!n=n_e-n_h$.
	In our case, the available $B$ likely remains below $B_H$ for DyPtBi, while for DyPdBi, this threshold may be reached as evidenced by clear maxima in $S_{xx}(B)$ at lower temperatures, which may also suggest that DyPtBi has a higher degree of carrier compensation, i.e., a smaller $\Delta\!n$ than DyPdBi.

	\subsection*{Correlation between longitudinal magnetothermopower and magnetoresistance}
	
\begin{figure}[h]
		\includegraphics[width=0.49\textwidth]{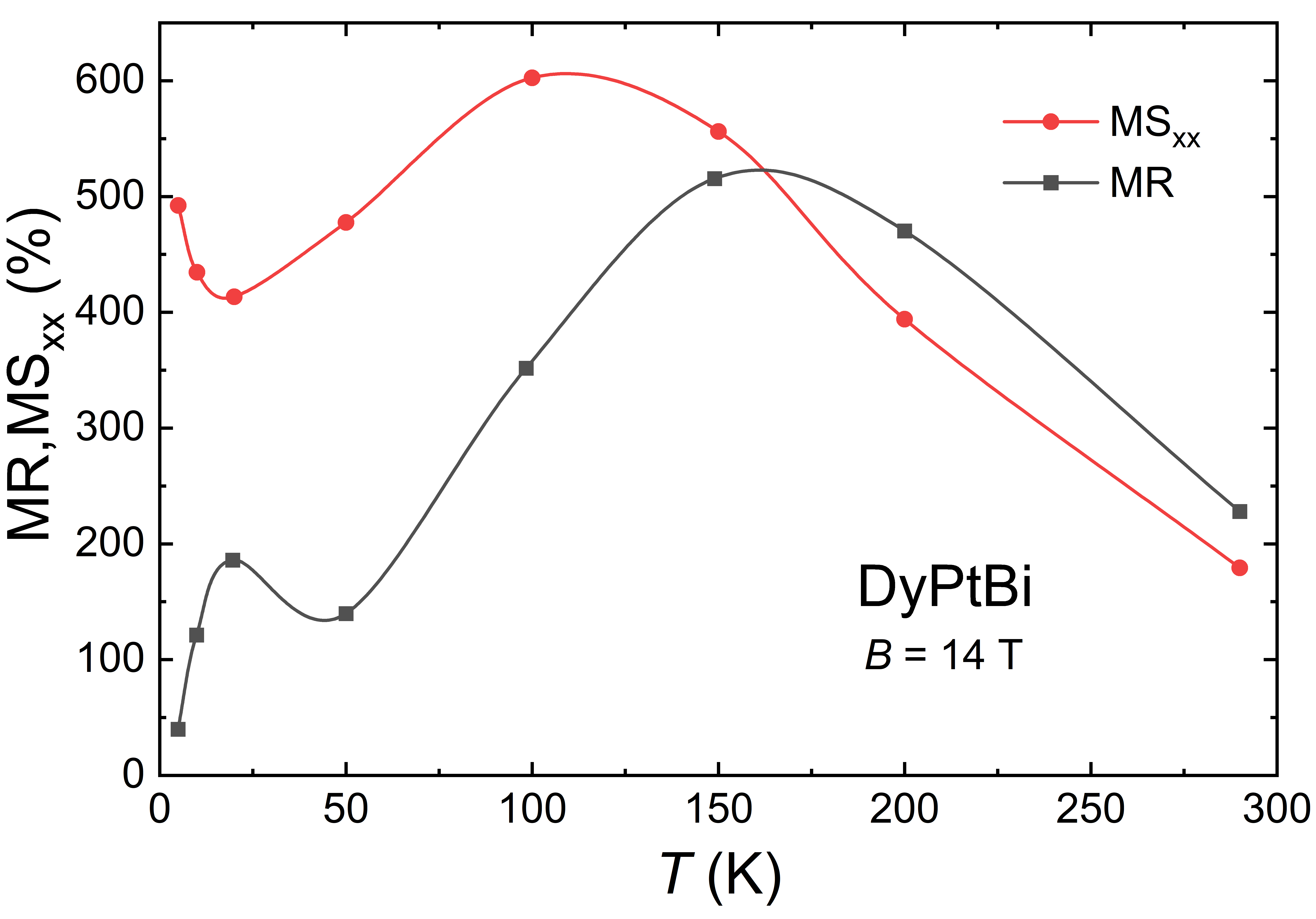}
		\includegraphics[width=0.49\textwidth]{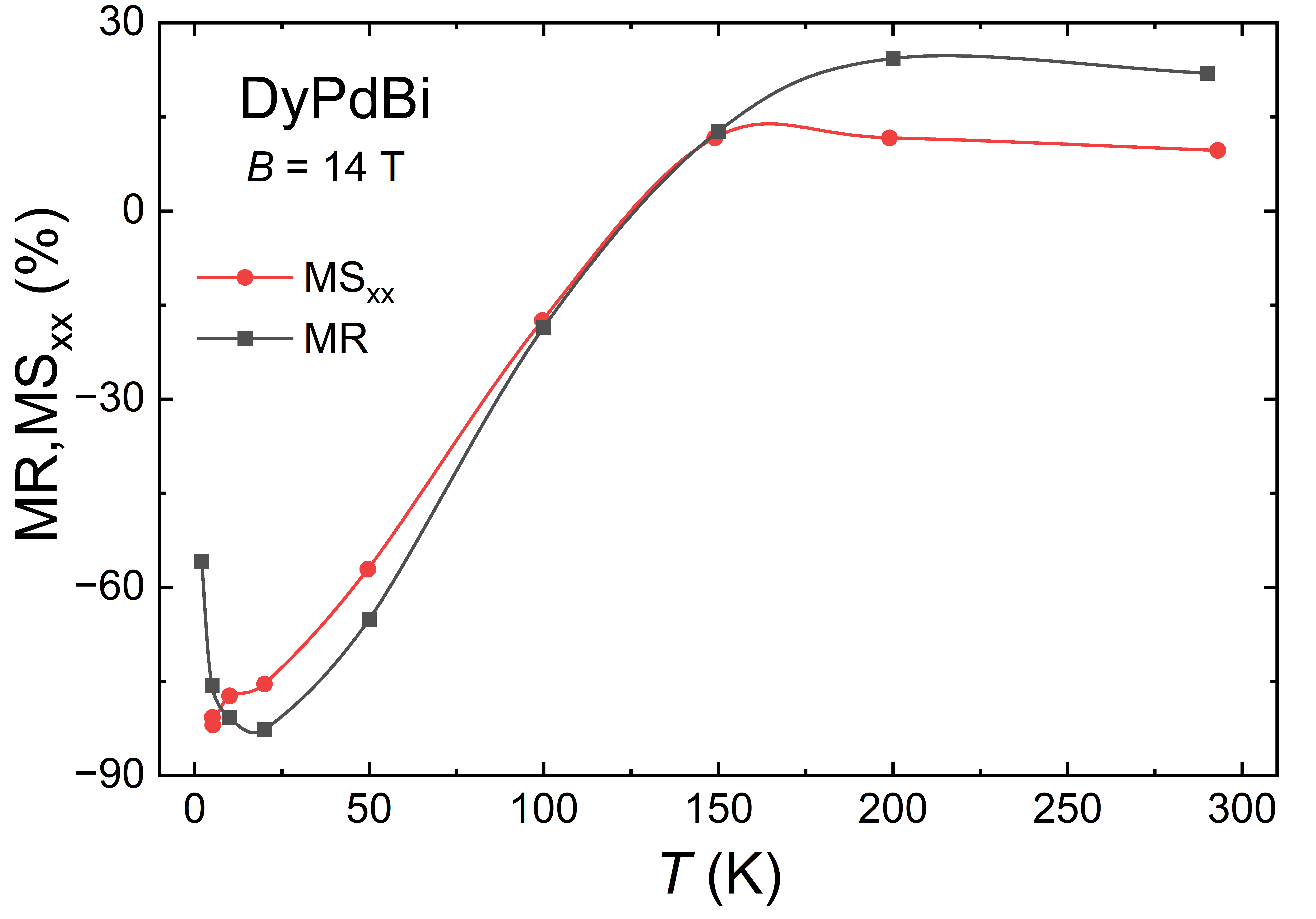}
		\caption{Temperature dependence of magnetoresistance and magneto-Seebeck thermopower measured in $B=14$\,T for DyPtBi (a) and DyPdBi (b).
			\label{MR_MS_vs_B}}
	\end{figure}

	In several recent studies, a correlation has been observed between magnetoresistance and longitudinal magneto-thermoelectric effect.\cite{Feng2022a1,Wang2022h1}
	In Fig.\,\ref{MR_MS_vs_B}a,b, the temperature dependence of magnetoresistance ($\rm{MR}=100\%\times[\rho(B)/\rho(0)-1]$) and magneto-$S_{xx}$ ($\rm{MS}_{xx}=100\%\times[S_{xx}(B)/S_{xx}(0)-1]$) are shown for $B=14$\,T.
	In the case of DyPdBi, both $\rm{MR}(T)$ and $\rm{MS}_{xx}(T)$ follow similar patterns across the temperature range from 20\,K to 300\,K. 
	In contrast, for DyPtBi both dependences exhibit similar overall behavior, but their characteristic maxima occur at different temperatures. 
	This suggests that the same mechanism may be responsible for both studied properties.   

	\subsection*{Magnetotransport}
	
	According to the Drude theory, MR($B$) in well-compensated semimetals is expected to be large and non-saturating even in strong magnetic fields, in stark contrast to the observations in usual metals, where MR increases quadratically with $B$ increasing in low magnetic fields and saturates at higher magnetic fields.\cite{Pippard2009}
	In turn, $\rho_{yx}(B)$ in such semimetals is expected to deviate from the linear field dependence characteristic of ordinary metal.
	
\begin{figure}[h]
		\includegraphics[width=0.49\textwidth]{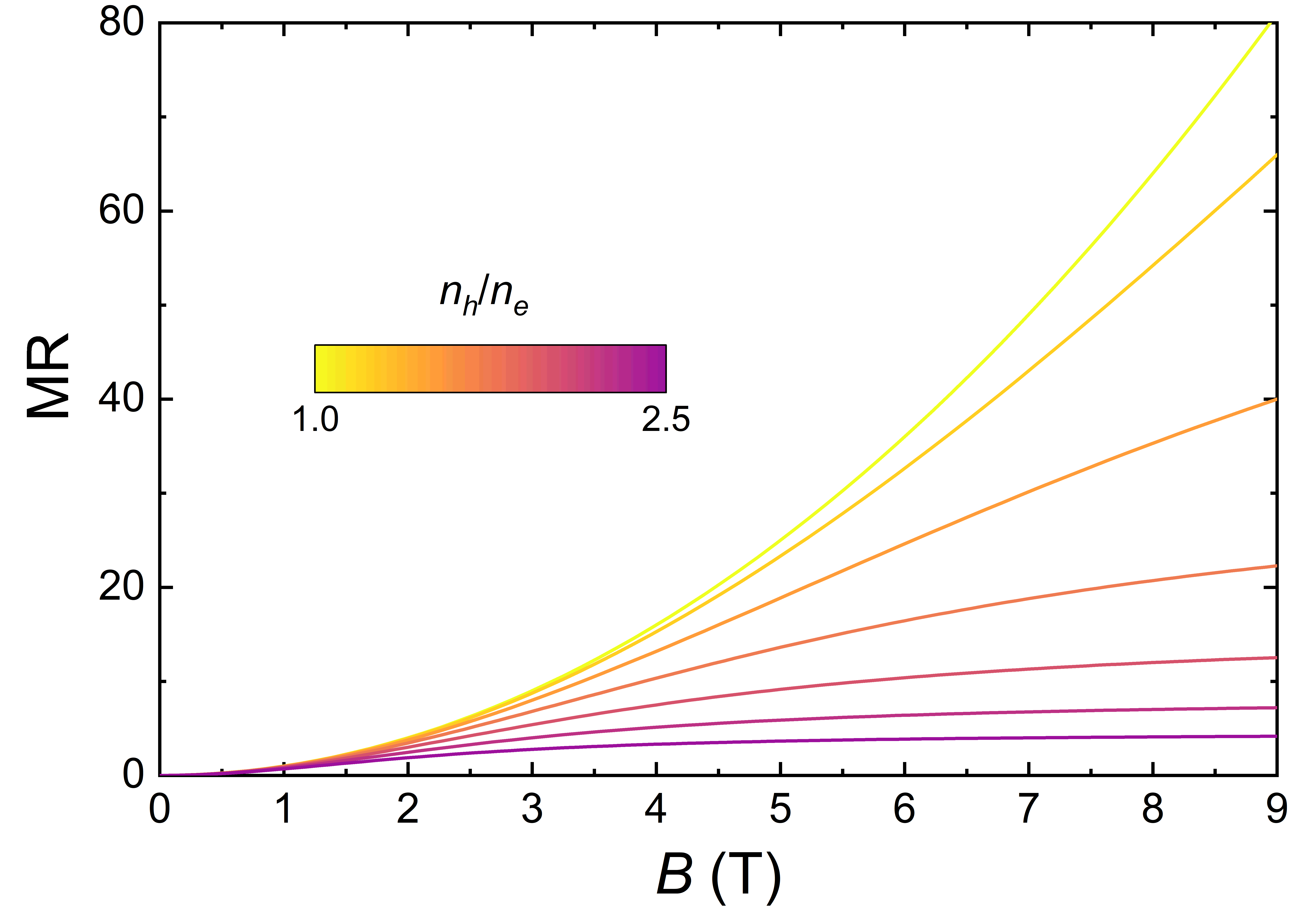}
		\caption{Simulated magnetic field dependence of magnetoresistance for different degrees of carrier compensation, based on the two-band Drude model (Eq.\,\ref{Drude}). The mobilities of hole and electron carriers were assumed to be equal to 1000\,cm$^2$/(Vs). 
			\label{MR_two-band}}
	\end{figure}

	The different behavior of MR($B$) in two compounds can be associated with the differences in electronic structures of the compounds studied (see Fig.\,1 and Electronic structure calculations subsection in main text). 
	The positive and non-saturated MR observed in DyPtBi can be attributed to the nearly-perfect carrier compensation. 
	In turn, the smaller degree of carrier compensation in DyPdBi leads to MR that is dominated by one of the mechanisms responsible for negative MR. 
	In our previous work,\cite{Pavlosiuk2019_1} we attributed negative MR seen in DyPdBi to a reduction of the spin-disorder scattering, while in a recent study of another half-Heulser material, TbPdBi, the authors proposed a different explanation for this phenomenon. 
	This is attributed to the high spin-polarisation of carriers caused by the ferromagnetic half-topological semimetal state.\cite{Zhu2023b1}
	In the region of weak $B$, MR of DyPdBi shows a rather sharp increase as $B$ increases. 
	In many previous works related to half-Heusler compounds,\cite{Pavlosiuk2016a1,Phys2019} this behavior was attributed to the weak anti-localization effect.  
	However, the alternative hypotheses are that this MR may originate from the classical magnetoresistance, as occurs in simple metals or from the semimetallic nature of this compound.\cite{Pippard2009} 
	When the degree of compensation is small, MR$(B)$ exhibits a behavior similar to that reported in simple metals (see Fig.\,\ref{MR_two-band}). 
	In general, the overall MR of DyPdBi can be understood as a competition between two contributions, negative and positive. 
	The former one is more pronounced at low temperatures and at high magnetic field, whereas the latter one dominates at higher temperatures and also at lower $T$ but in the weak-$B$ region.

	One may ask why DyPtBi does not exhibit negative magnetoresistance, given that it possesses the same magnetic structure with propagation vector ${\bm{k}}=(1/2, 1/2, 1/2)$ as DyPdBi\cite{Pavlosiuk2018,Zhang2020k1} and magnetic field dependence of magnetization for both compounds are similar in magnitude. 
	We can also assume that the magnitude of the negative MR arising from the reduced spin-dependent scattering is likely comparable in both compounds. 
	This hypothesis is supported by the fact that DyPtBi and DyPdBi are similar not only magnetically (the same magnetic structure, similar values of $T_N$, comparable magnetization behavior) but also electronically (comparable electrical resistivity and similar behavior of $\rho(T)$) in zero magnetic field.
	Thus, the key factor differentiating their overall magnetoresistance is the degree of carrier compensation.  
	
\begin{figure}[h]
		\includegraphics[width=0.49\textwidth]{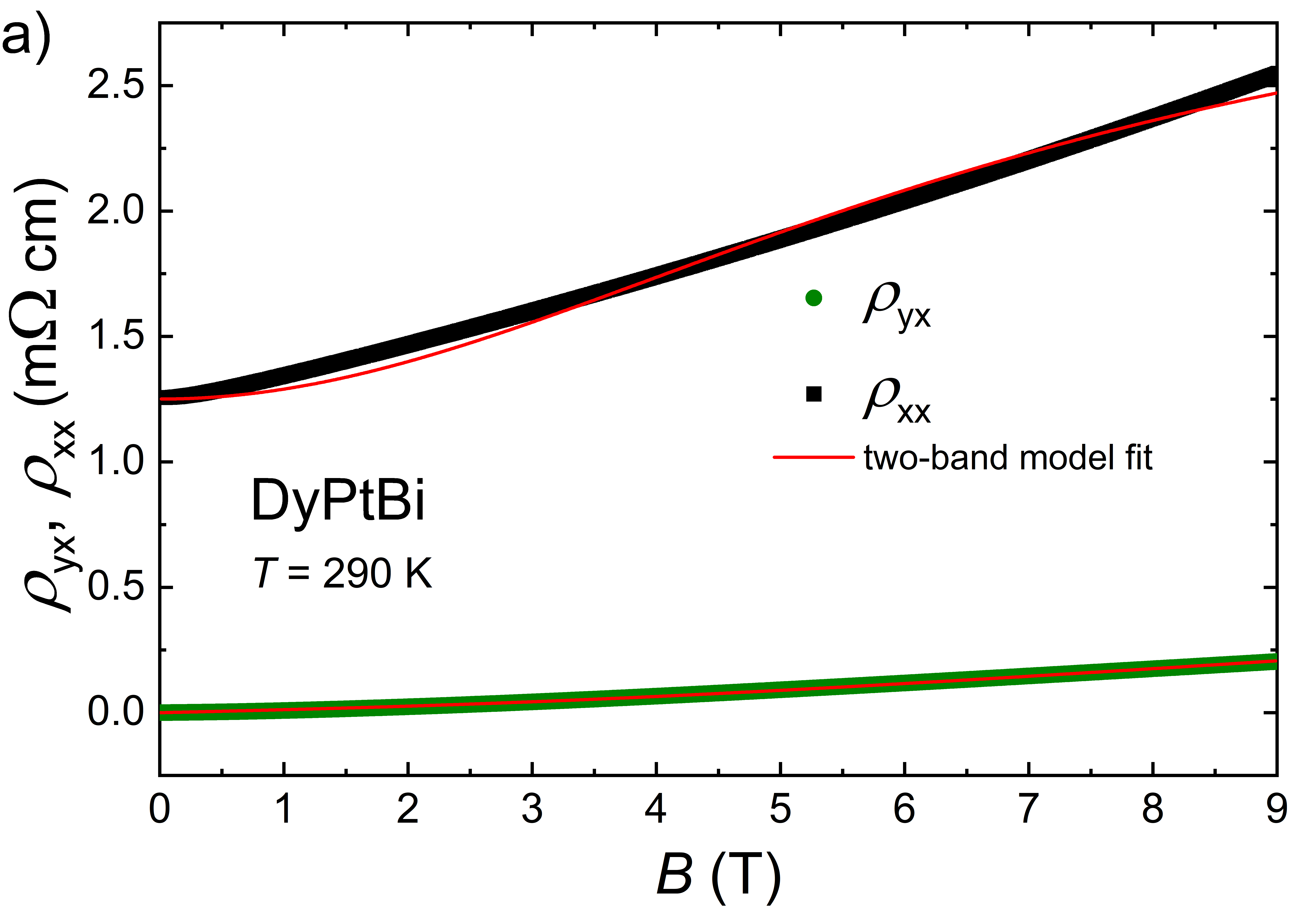}
		\includegraphics[width=0.49\textwidth]{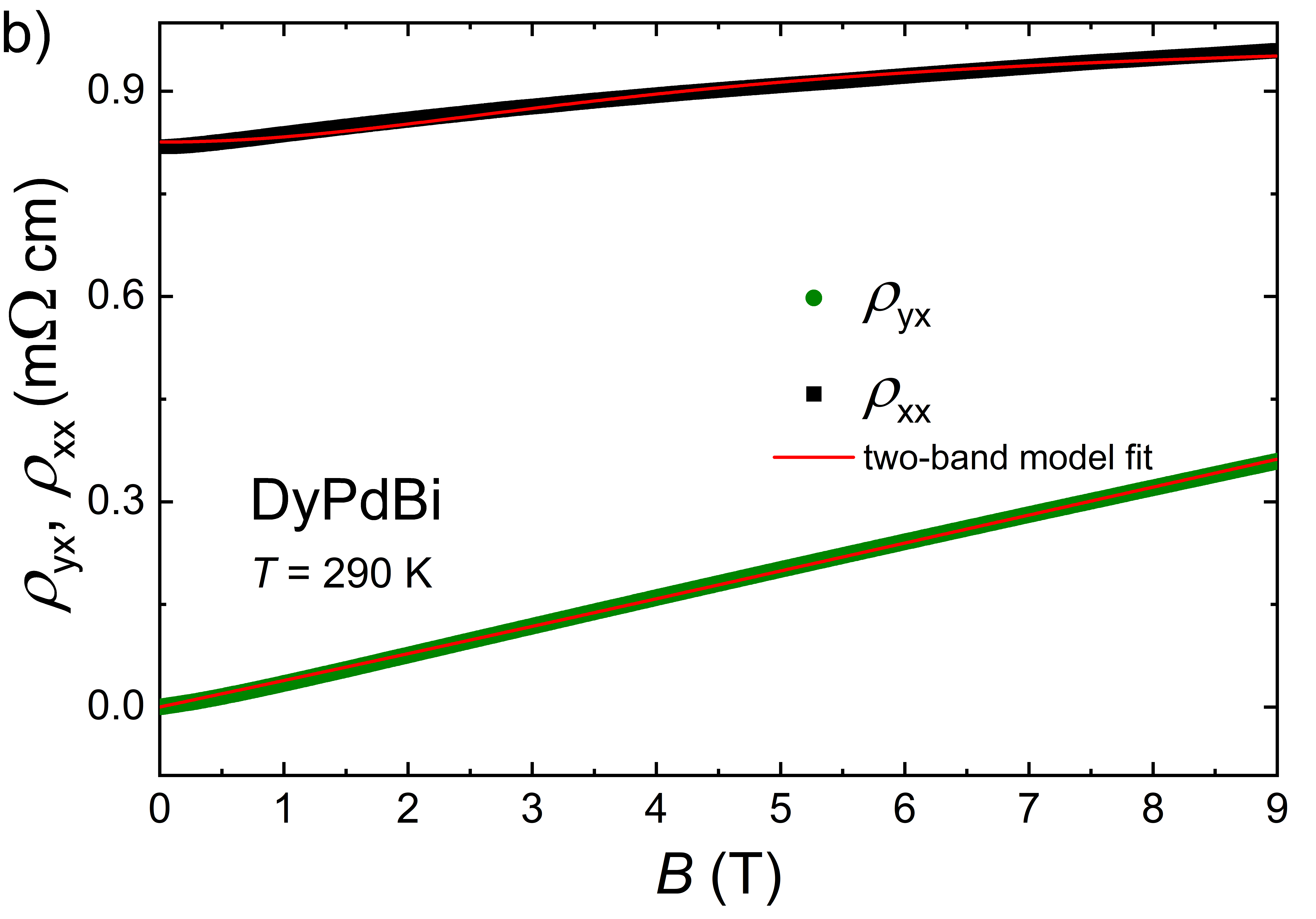}
		\caption{Global fitting of $\rho_{yx}(B)$ and $\rho_{xx}(B)$ measured at $T=290$\,K for DyPtBi (a) and DyPdBi (b) using the Drude two-band model (Eq.\,\ref{Drude}). The fitted curves are shown as red solid lines.
			\label{rho_two_band}}
	\end{figure}

	Fig.\,\ref{rho_two_band} presents the results of fitting the two-band Drude model (Eq.\,\ref{Drude}):
	
	\begin{equation}
		\begin{matrix*}[l]
			\rho_{yx}(B) = \frac{B}{e} \cdot \frac{(n_h \mu_h^2 - n_e \mu_e^2) + (n_h - n_e)\mu_h^2 \mu_e^2 B^2}{(n_h \mu_h + n_e \mu_e)^2 + (n_h - n_e)^2 \mu_h^2 \mu_e^2 B^2}, \\\\
			\rho_{xx}(B) = \frac{1}{e} \cdot \frac{(n_h \mu_h + n_e \mu_e) + (n_e \mu_e \mu_h^2 + n_h \mu_h \mu_e^2) B^2}{(n_h \mu_h + n_e \mu_e)^2 + (n_h - n_e)^2 \mu_h^2 \mu_e^2 B^2}
		\end{matrix*}
		\label{Drude}
	\end{equation} 
	
	to magnetic field dependence of Hall resistivity ($\rho_{yx}$) and longitudinal resistivity ($\rho_{xx}$) measured at $T=290$\,K for DyPtBi and DyPdBi. 
	For each compounds, both datasets were fitted simultaneously using a single set of fitting parameters (global fitting), where $n_e$ and $n_h$ are the electron and hole carrier concentrations, respectively, and $\mu_e$ and $\mu_h$ stand for their corresponding mobilities. 
	The least-square fitting yields the following values of parameters: (i) for DyPtBi: $n_e=6.3\!\times\!10^{24}$\,cm$^{-3}$, $n_h=2.7\!\times\!10^{25}$\,cm$^{-3}$, $\mu_e=2300$\,cm$^2$/(Vs), $\mu_e=2324$\,cm$^2$/(Vs)
	(ii) for DyPdBi $n_e=4.8\!\times\!10^{23}$\,cm$^{-3}$, $n_h=1.6\!\times\!10^{25}$\,cm$^{-3}$, $\mu_e=2345$\,cm$^2$/(Vs), $\mu_h=8440$\,cm$^2$/(Vs). 
	These parameters indicate that DyPtBi has a higher degree of carrier compensation than DyPdBi.

\end{document}